\documentclass{article}

\PassOptionsToPackage{numbers,sort&compress}{natbib}

\usepackage[preprint]{neurips_2025}

\usepackage[utf8]{inputenc}
\usepackage[T1]{fontenc}    
\usepackage{hyperref} 
\usepackage{url}     
\usepackage{booktabs}    
\usepackage{amsfonts}    
\usepackage{nicefrac}  
\usepackage{microtype}    
\usepackage{graphicx}
\usepackage{enumitem}

\usepackage{booktabs}
\usepackage{multirow}
\usepackage[table,xcdraw]{xcolor}

\usepackage{wrapfig}
\usepackage{float}
\usepackage{array, booktabs}
\usepackage{amsmath} 
\usepackage{afterpage}
\usepackage{graphicx}
\usepackage{caption}
\usepackage{adjustbox}
\usepackage{calc}
\usepackage{setspace}
\usepackage{placeins}

\usepackage{titletoc}
\usepackage{varwidth}
\usepackage{graphicx} 
\usepackage{lineno}

\usepackage{bm}

\usepackage{amsfonts, amssymb}
\definecolor{mycolor}{HTML}{6A9955}

\usepackage{algorithm}
\usepackage{algorithmicx}
\usepackage{algpseudocode}
\floatname{algorithm}{Algorithm}
\usepackage{setspace}

\usepackage{wrapfig}
\usepackage{listings}
\usepackage{xcolor}
\title{iDSE: Navigating Design Space Exploration in High-Level Synthesis Using LLMs}

\author{
  Runkai Li\thanks{Equal contribution.}\\
  Southeast University\\
  \And
  Jia Xiong\footnotemark[1]\\
  Southeast University \\
  \And
  Xi Wang\thanks{Corresponding author: Xi Wang (xi.wang@seu.edu.cn).}\\
  Southeast University \\
}

\begin{document}

\maketitle

\begin{abstract}

High-Level Synthesis (HLS) serves as an agile hardware development tool that streamlines the circuit design by abstracting the register transfer level into behavioral descriptions, while allowing designers to customize the generated microarchitectures through optimization directives.
However, the combinatorial explosion of possible directive configurations yields an intractable design space.
Traditional design space exploration (DSE) methods, despite adopting heuristics or constructing predictive models to accelerate Pareto-optimal design acquisition, still suffer from prohibitive exploration costs and suboptimal results. 
Addressing these concerns, we introduce iDSE, the first LLM-aided DSE framework that leverages HLS design quality perception to  effectively navigate the design space. 
iDSE intelligently prunes the design space to guide LLMs in calibrating representative initial sampling designs, expediting convergence toward the Pareto front. By exploiting the convergent and divergent thinking patterns inherent in LLMs for hardware optimization, iDSE achieves multi-path refinement of the design quality and diversity.
Extensive experiments demonstrate that iDSE outperforms heuristic-based DSE methods by 5.1$\times$$\sim$16.6$\times$ in proximity to the reference Pareto front, matching NSGA-II with only 4.6\% of the explored designs.
Our work demonstrates the transformative potential of LLMs in scalable and efficient HLS design optimization, 
offering new insights into multiobjective optimization challenges.

\end{abstract}

\section{Introduction}
\label{sec:Introduction}
\begin{figure*}[t]
	\centering 
        \includegraphics[width=14cm]{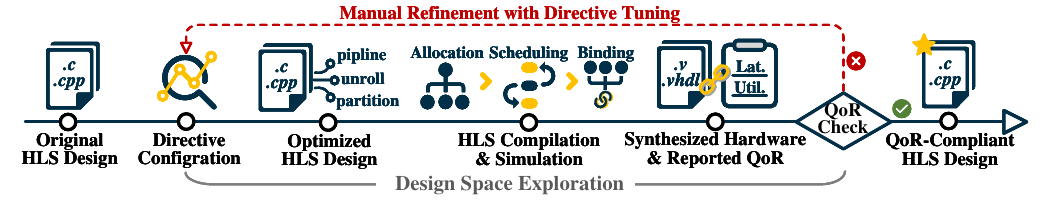} 
        \vspace{-13pt}
	\caption{Time-consuming manual directive configuration tuning based on QoR reported in HLS.}
	\label{workflow1}
    \vspace{-10pt}
\end{figure*}

Transistor scaling has driven exponential performance improvements in modern circuits, yet it remains incapable of accommodating the escalating computational demands in fields such as neural networks \cite{DSA_ML_1, DSA_ML_2, DSA_ML_3, DSA_ML_4}, computer vision \cite{DSA_CV_1, DSA_CV_2}, robotics \cite{DSA_Robotics_1, DSA_Robotics_2}, and genome sequence analysis \cite{DSA_GSA_1, DSA_GSA_2, DSA_GSA_3, DSA_GSA_4}.
The development of domain-specific accelerators (DSAs) tailored to specific workloads has shown notable growth \cite{dark_silicon, Democratizing}.
However, the protracted chip design and verification cycles impede the swift iteration of DSAs required to keep pace with the rapidly evolving dedicated application requirements.
High-Level Synthesis (HLS) abstracts hardware description languages (HDL) into behavioral representation, accelerating FPGA-based prototyping with comparable circuit performance and improved hardware development efficiency \cite{HLS_review1, HLS_review2}. 
Furthermore, most vendor HLS tools incorporate optimization directives to customize microarchitectures, synthesized from high-level programming languages, to align design requirements with quality of results (QoR). 
However, the extensive permutations of directive configurations constitute an expansive design space, rendering the identification of Pareto-optimal designs challenging and expertise-intensive, as illustrated in Figure \ref{workflow1}.

To capitalize the reconfigurability and scalability advantages of HLS and advance hardware optimization, recent research has focused on automated and efficient design space exploration (DSE) methods aimed at reducing manual intervention.
However, the extensive combination of directives exponentially expands the design space. 
Furthermore, the time-consuming performance evaluation required by HLS tools renders exhaustive traversal impractical.
Traditional DSE methods typically treat HLS as a black-box optimization problem, relying heavily on heuristics that necessitate multiple iterations to approximate Pareto-optimal solutions. This approach often results in prolonged exploration periods and inadequate coverage of optimal designs \cite{DSE_SA, DSE_GA, DSE_ACO, DSE_Bayesian, DSE_lattice, DSE_S2FA}.
In response, machine learning (ML) and deep learning (DL) techniques have revisited DSE by leveraging predictive models to surrogate expensive HLS evaluations.
These approaches significantly improve efficiency by enabling the evaluation of substantially more designs \cite{HLSYN, DSE_GNNDSE, DSE_HARP, DSE_IRONMAN, DSE_COMBA, DSE_Lin, DSE_GNNHLS, DSE_Compare, hgbo}.
However, model-based methods typically involve significant training overhead and exhibit limited generalization across unseen workloads, HLS environments, or varied hardware constraints, potentially compromising the optimality of explored designs.

Recently, Large Language Models (LLMs) have attracted considerable attention for their exceptional proficiency in natural language processing and code generation. However, their effectiveness in specialized applications is often hampered by limited domain-specific data. Previous efforts have harnessed the code generation and debugging capabilities of LLMs to streamline chip design and verification workflows, highlighting their potential to improve efficiency within electronic design automation (EDA) \cite{LLM_ChatCPU, LLM_MEIC, LLM_ChatEDA, LLM_RTLFixer, LLM_GPT4AIGChip}. 
Despite these advancements, existing approaches have not sufficiently exploited LLMs to achieve a high degree of hardware optimization for high-quality circuits required by emerging applications. 
Recent research has demonstrated considerable promise in employing LLMs for gradient-free black-box optimization tasks \cite{LLM_opt1, LLM_opt2, LLM_opt3, LLM_opt4, LLM_EC_Review}.
Such approaches leverage the natural language reasoning capabilities of LLMs, enabling iterative and informed exploration of optimization trajectories through linguistic prompting. 
HLS abstracts hardware designs to the algorithmic level while ensuring functional equivalence, providing an opportunity to empower hardware optimization by directing LLMs towards the design structure, thus reducing hardware specification semantics. 
Furthermore, leveraging the inherent domain expertise and reasoning capabilities in LLMs, it becomes feasible to efficiently allocate operational parallelism in HLS-generated hardware.

Building on this insight, we propose \textbf{iDSE}, a novel LLM-navigated DSE framework that reduces time costs and expertise barriers for HLS design optimization.
This framework automates the end-to-end process by extracting HLS design features and pruning invalid designs, thus effectively distilling the design space.
Furthermore, iDSE incorporates a warm-start mechanism, which initializes exploration with representative seed designs to accelerate convergence to the Pareto front. By employing operators inspired by evolutionary algorithms to spawn refined design candidates, iDSE capitalizes on prior knowledge to analyze design bottlenecks, identify optimization opportunities and refine designs. Quantitative metrics highlight that our framework shapes a broader and more concave Pareto front, providing a new perspective for promoting the customization of DSAs.

The main contributions of this paper are as follows:

\begin{itemize}[leftmargin=*]

\item We introduce \textbf{iDSE}, the \textbf{first} end-to-end design space exploration framework integrating optimization trajectory awareness with prior expertise injection. This presents a compelling direction for conquering the intricate challenges of automated hardware optimization.

\item We propose a scalable \textbf{Feature-Driven Pruning} approach to significantly expedite the DSE iterations by constructing a compact yet expressive HLS design space. 

\item We reimagine DSE initialization through the LLM-guided \textbf{Seed Directive Generation} methodology. This approach enables warm-starts that rapidly converge toward comprehensive and concave Pareto front profiling by improving the quality and diversity of the initial sampling designs.

\item We introduce a novel \textbf{QoR-Aware Adaptive Optimization} system that exploits the convergent and divergent thinking capabilities of LLMs to navigate multi-path HLS design optimization. By perceiving QoR feedback, our system transcends existing bottlenecks and escapes local optima.

\item Our extensive experiments across diverse HLS benchmarks demonstrate \textbf{5.1\bm{$\times$}}\bm{$\sim$}\textbf{16.6\bm{$\times$}} improvement in explored Pareto front quality over heuristic-based DSE methods, with up to \textbf{25.1\bm{$\times$}} higher exploration efficiency. This work reveals the unique potential of LLM-aided hardware optimization.

\end{itemize}

\section{Background \& Related Work}
\label{sec:Background}
\begin{figure*}[t] 
	\centering 
        \includegraphics[width=14cm]{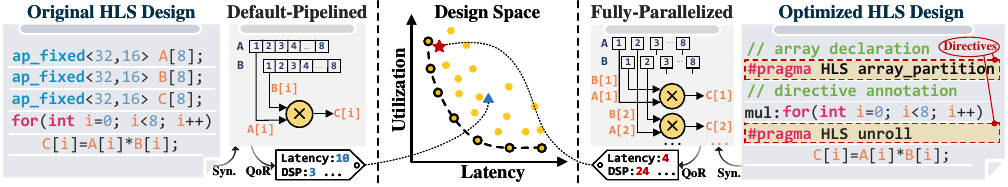} 
        \vspace{-13pt}
	\caption{Example of customizing synthesized hardware with HLS optimization directives. \textbf{Left:} Default loop \textit{read-compute-write} pipelining structure without specifying any directives. \textbf{Middle:} QoR mapping across the design space. \textbf{Right:} Fully parallelized memory access and loop operations.}
	\label{workflow2}
    \vspace{-10pt}
\end{figure*}

\textbf{High-Level Synthesis Design Space Exploration.} HLS employs optimization directives to transform high-level description languages (C/C++/System C) into specific microarchitectures through hardware resource allocation, and operation scheduling/binding. While designers typically prioritize Pareto-optimal solutions that satisfy certain constraints, the exponentially growing design space compiled by the Cartesian product of directives renders brute-force traversal computationally infeasible, particularly given the time-consuming quality of results (QoR) evaluation for each configuration. Figure \ref{workflow2} illustrates a toy example for implementing and optimizing a vector Hadamard product in HLS. 
Although this HLS design contains only one loop and three arrays, considering the loop and memory access parallelization directives supported in this paper, where the factors for loop unrolling and memory partitioning are divisors of loop and array boundaries, we still obtain \textbf{1.58M valid} designs composed of various combinations and parameters of directives. Exhaustively evaluating all designs would require approximately \textbf{3 years}, assuming only one minute per configuration evaluation.

Meta-heuristic \cite{DSE_SA, DSE_GA, DSE_ACO, DSE_Bayesian, DSE_CEM} and dedicated heuristic \cite{AutoDSE, DSE_S2FA, DSE_Dedicated_1, DSE_Dedicated_2} DSE methods treat the HLS tool as a black box, leveraging hardware optimization characteristics to efficiently approximate Pareto-optimal solutions. While exhaustive searches are avoided, these methods still necessitate multiple invocations of the HLS tool to guide optimization trajectories. Moreover, their effectiveness remains constrained by initial sampling quality, limiting exploration within a confined design space under restricted search budgets. 
The introduction of ML and DL methods has driven the construction of analysis models for QoR prediction \cite{DSE_COMBA, DSE_Lin, DSE_Sisphus, DSE_NLP}.
By providing surrogate models for HLS tool evaluations, these methods enable performance and resource utilization predictions for the synthesized hardware. 
Graph neural networks (GNNs) embed program nodes and directive configurations have notably improved prediction accuracy \cite{DSE_GNNDSE, DSE_HARP, DSE_IRONMAN, DSE_GNNBENCH}. 
However, ML/DL-based approaches demand substantial training and deployment costs, exhibit limited generalization across different applications and versions of vendor HLS tools. Additionally, these methods often focus on optimizing a single performance metric, thereby overlooking the full spectrum of optimization requirements.

\textbf{LLM-Aided Hardware Design.} Integration of LLMs has catalyzed the evolution of EDA tools \cite{LLM_ChatCPU, LLM_MEIC, LLM_ChatEDA, LLM_RTLFixer, LLM_LLM4EDA, LLM_BetterV, LLM_LiK, LLM_UVLLM, LLM_VGV, LLM_VeriGen, LLM_ADO, LLM_LEDRO, LLM_veridebug, LLM_insights}.
However, current efforts struggle to keep pace with the rapidly evolving application requirements, and achieving modern high-performance computing architectures remains challenging, necessitating breakthroughs in LLM-aided hardware optimization. Previous research on LLMs applying to HLS has examined their capability to insert directives into source code, utilizing knowledge-augmented technology to bridge the expertise gap in HLS \cite{LLM_HLS_1, LLM_HLS_4, LLM_HLS_5, LLM_HLS_6, LLMDSE}. However, the results have been underwhelming, with code transformations frequently introducing synthesis errors. Furthermore, existing studies primarily target performance-optimal designs, identifying HLS designs across the Pareto front to accommodate diverse optimization preferences remains nascent \cite{LLM_HLS_2, LLM_HLS_3}. This research gap obscures the practical implementation of LLMs in hardware optimization contexts.

\textbf{Machine Learning in Multiobjective Optimization.} 
Different from conventional multiobjective optimization methods \cite{Pareto_1, Pareto_2, Pareto_3, PSL_1, PSL_2, PSL_3, PSL_4, PSL_5}, 
recent research has demonstrated significant improvements in heuristic optimization \cite{LLM_EoH, LLM_FunSearch, LLM_EC_Review, LLM_EC_MCTS, LLM_EC_MEoH, LLM_EC_Reevo} and neural architecture search \cite{EvoPrompting, GPT-NAS, LLM_EC_llmatic, LLM_EC_graph, LLM_EC_LLM, LLM_EC_design} by embedding LLMs into evolutionary algorithms (EA). This approach outperforms manual tuning and traditional automated approaches, highlighting the promise of LLM-driven multiobjective optimization. 
Some research has employed LLM-enhanced EA operators in conjunction with GNN-based predictive models for DSE \cite{DSE_LLM_EA}. 
However, substituting actual evaluation with regression models inevitably compromises performance. Meanwhile, existing LLM implementations of EA often yield suboptimal results due to inadequate reflection on optimization trajectories and a lack of task-specific guidance for DSE.

\section{Preliminary \& Problem Formulation}
\label{sec:Problem Formulation}
Balancing computation and memory access parallelism under hardware resource constraints to achieve satisfactory circuit performance is a delicate process. Since performance improvements and resource consumption are often contradictory, DSE constitutes a multiobjective optimization problem.
We focus on two primary objectives in DSE: \textbf{execution latency} and \textbf{resource utilization}. For an HLS design \(\lambda(\varphi)\), \(\varphi\) represents the inserted optimization directives, this work emphasizes three directives, \texttt{PIPELINE} (\(\mathcal{LP} \)), \texttt{UNROLL} (\(\mathcal{LU}\)), \texttt{ARRAY\_PARTITION} (\(\mathcal{AP}\)), 
which control loop execution and memory access parallelism. 
Define \(\varphi\) as a feature vector
\(
\varphi = \left[ \mathcal{LP}_i, \mathcal{LU}_j, \mathcal{AP}_{k,d} \right]
\)
, where \(\mathcal{LP}_i\) is a boolean used to enable loop \(i\) pipelining, \(\mathcal{LU}_j\) is the unroll factor for loop \(j\), and \(\mathcal{AP}_{k,d}\) is the partition type and factor for array \(k\) along dimension \(d\). Under vendor HLS tool \(\mathcal{H}\), we analyze the QoR of the explored design using latency \(Lat(\mathcal{H}, \lambda(\varphi))\) and resource utilization \(Util(\mathcal{H}, \lambda(\varphi))\).

\textbf{Definition 1} (Multiobjective Optimization of DSE). The multiobjective DSE task is defined as:  

\vspace{-7pt}

\begin{equation}\label{eq1} 
\lambda(\varphi^*) = 
\mathop{\arg\min}_{\varphi \in \Phi \subset \mathbb{Z}^n} \left[Lat(\mathcal{H}, 
\lambda(\varphi)), Util(\mathcal{H}, \lambda(\varphi)) \right]
\end{equation}

\vspace{-5pt}

We aim to optimize both objectives without significantly compromising either. The goal of DSE is to rapidly and accurately search the design space $\Phi$ for Pareto-optimal designs.

\textbf{Definition 2} (Pareto-Optimal Designs). The explored designs \(\lambda(\varphi^*)\) and \(\lambda(\varphi^a)\) if:  

\vspace{-8pt}

\begin{equation}\label{eq2} 
Lat(\mathcal{H}, \lambda(\varphi^*)) \leq Lat(\mathcal{H}, \lambda(\varphi^a)), \quad Util(\mathcal{H}, \lambda(\varphi^*)) \leq Util(\mathcal{H}, \lambda(\varphi^a))
\end{equation}

\vspace{-3pt}

We call it \(\lambda(\varphi^*)\) dominates \(\lambda(\varphi^a)\). If no other \(\varphi \in \Phi\) dominates \(\lambda(\varphi^*)\), then \(\lambda(\varphi^*)\) is called a Pareto-optimal design. All such designs form the Pareto front. 

\textbf{Definition 3} (Effectiveness of DSE). We employ the average distance to reference set (ADRS) metric to quantify the gap \(d(\cdot)\) between an explored Pareto front \({P}_E\) and a reference Pareto front \({P}_R\) \cite{DSE_Review}:  

\vspace{-5pt}

\begin{equation}\label{eq3} 
ADRS({P}_E, {P}_R) = \frac{1}{|{P}_R|} \sum_{\lambda(\varphi^\gamma) \in {P}_R} \min_{\lambda(\varphi^\omega) \in {P}_E} d(\lambda(\varphi^\gamma), \lambda(\varphi^\omega))
\end{equation}

\vspace{-5pt}

A small ADRS indicates that the explored designs more effectively approximate the entire reference Pareto front. 
Achieving a small ADRS within constrained search budgets demonstrates the performance and efficiency of the proposed DSE method.

\section{iDSE Design \& Philosophy}
\label{sec:Methodology}
\begin{figure*}[t] 
	\centering 
        \includegraphics[width=14cm]{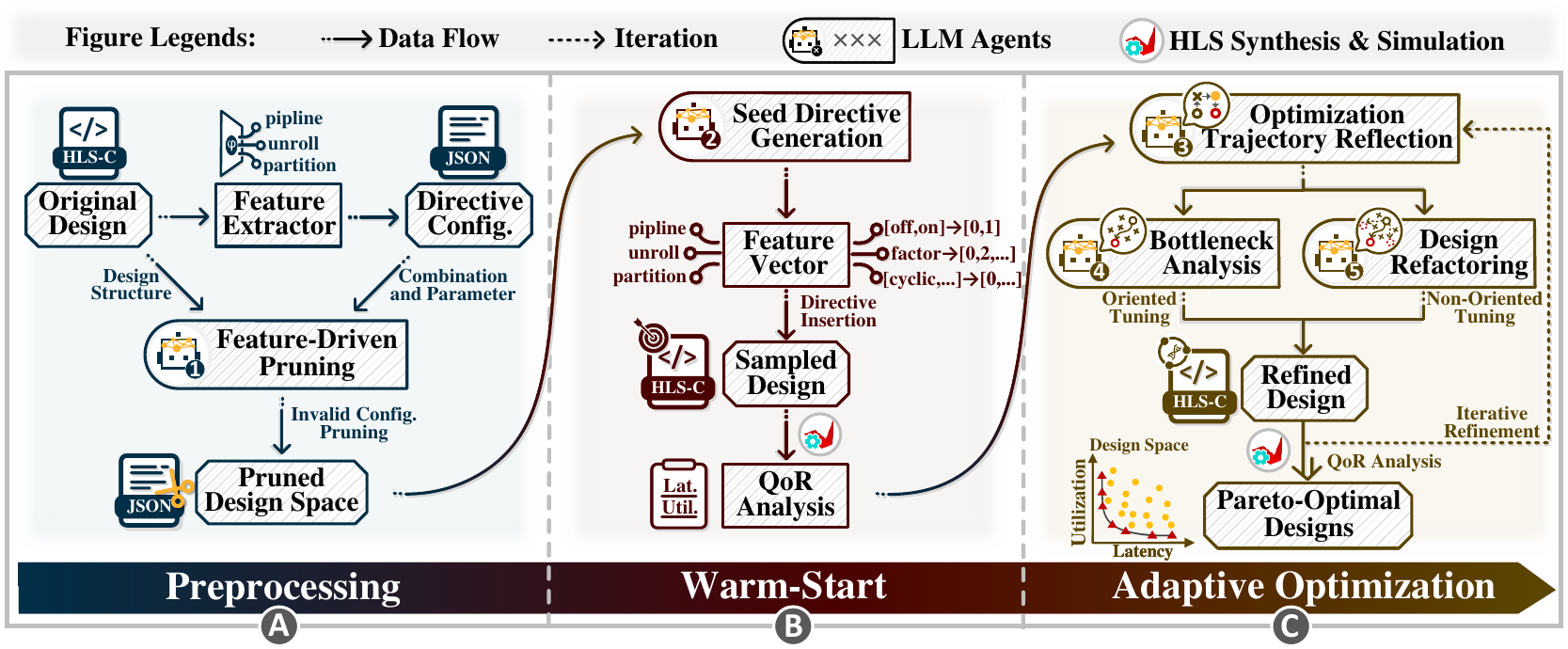} 
        \vspace{-13pt}
	\caption{iDSE Workflow}
	\label{main workflow}
    \vspace{-15pt}
\end{figure*}

Our framework, \textbf{iDSE}, automates the exploration of design spaces to optimize HLS designs with LLM as the backbone, effectively identifying Pareto-optimal designs while balancing competing objectives. The workflow of iDSE, depicted in Figure \ref{main workflow}, unfolds in three distinct stages:

\textbf{\textit{1) Preprocessing.}}
iDSE first extracts a unified design space from the provided HLS design and employs specific pruning strategies to eliminate invalid directive configurations that may result in ineffective or inefficient synthesis. This step defines a feasible domain and narrows the parameter ranges to be searched, thereby improving the efficiency of approximating the Pareto front.

\textbf{\textit{2) Warm-Start.}}
iDSE then leverages the LLM to reproduce the learned insights to identify directive configurations with optimization potential from the expansive design space, providing high-quality and discrete sampling designs to warm-start subsequent optimization.

\textbf{\textit{3) Adaptive Optimization.}}
To further approximate the Pareto front, iDSE performs a bottleneck analysis on initial sampling designs, guided by empirical optimization knowledge, which prompts LLM to conduct oriented reasoning for localized design refinement. Furthermore, iDSE exploits the divergent thinking of LLMs to escape limitations of existing optimization strategies and propose novel optimization directions, thus expanding the exploration scope and improving design diversity.

\subsection{Preprocessing: Directive Configuration Extraction and Design Space Pruning}

\vspace{-1pt}

The allocation of loop and memory access parallelism constitutes a critical bottleneck in hardware optimization, significantly affecting both performance and resource utilization of synthesized hardware designs. \textit{Feature Extractor} parses the design structure to extract key structural metadata (e.g., array dimensions, loop trip counts) that determine feasible parallelism degrees for optimization directive combinations and parameters. However, for large design footprints, this may introduce aggressive parallelism that makes the generated optimization strategy infeasible.

Building upon these extracted features, we propose the \textbf{Feature-Driven Pruning} approach, which eliminates invalid directive configurations that may lead to excessive synthesis time or even failure, thus effectively compressing the design space.
This approach prompts the LLM to adopt specific pruning strategies that intelligently identify and remove aggressive parallelism and invalid directive combinations.
For example, when examining nested loops with multilayer structures or large inner loop trip counts, judiciously disabling the pipeline option for the outer loop effectively eliminates about half of the invalid designs (pruning strategies are detailed in Appendix \ref{app:prune}).
Furthermore, we provide a prompt interface that allows designers to customize their pruning strategies according to specific hardware resource tolerance.
Importantly, this method preserves potential Pareto-optimal designs, which would not be extensively missed due to aggressive pruning strategies.

\vspace{-1pt}

\subsection{Warm-Start: LLM-Guided Design Space Exploration Initialization}
\label{warmstart}

\vspace{-1pt}

LLMs can effectively leverage prior knowledge of hardware optimization by emulating the reasoning patterns of seasoned engineers, as similar directives embedded in analogous design structures typically yield comparable QoR \cite{DSE_transfer_1, DSE_transfer_2}. 
We propose an LLM-guided \textbf{Seed Directive Generation} mechanism, which first generates a formatted feature vector of seed directive configuration, and then automatically creates a Tcl script to embed directives into the original HLS design. 
Meanwhile, this structured directive configuration enables seamless integration with heuristic-based DSE methods.

To ensure high-quality initial sampling designs, we establish specific design principles to guide the LLM in generating directive configurations that can be executed efficiently. This approach prevents the LLM from simply refluxing trivial solutions drawn from its pretraining data (detailed prompt design available in Appendix \ref{app:Prompt}). 
Our initialization strategy begins by defining the sampling task with universal optimization objectives: prioritizing performance, prioritizing resource utilization, and balancing both considerations. By providing the LLM with the structured directive configurations derived from the \textit{preprocessing} stage, we enable it to generate a prescribed number of diverse directive combinations that cover distinct regions of the design space. 
An automation tool then parses the execution latency and resource utilization metrics from the HLS synthesis report.

\vspace{-1pt}

\subsection{Adaptive Optimization: Multi-Path Directive Tuning with QoR Perception}

\vspace{-1pt}

\begin{algorithm}[t]
\small
\caption{Adaptive Optimization for HLS DSE}\label{algorithm}
\begin{algorithmic}[1] 
\Require LLM \({\pi_\theta}\), pruned design space $\Phi$, directive configuration $\varphi$, HLS design \(\lambda(\varphi)\), initial sampling size \(N_0\), maximum number of generations \(I_{max}\), adaptive population size \(P_i\), vendor HLS tool $\mathcal{H}$,
 \Ensure Pareto-optimal designs ${\lambda}(\varphi^*)$

 \setstretch{1.2}
 
 \State Initialize $P_0$ with $\varphi_{init} \leftarrow$ \textsc{WarmStart}(
${\pi_\theta }, \Phi, N_0, \lambda(\varphi)$
 ) 
\hfill \footnotesize $\triangleright$ Section \ref{warmstart}

\State Evaluate quality of results $\mathcal{Q}$ of initial sampling HLS designs $\lambda(\varphi_{init})$ using $\mathcal{H}$
 
\For{$i \leftarrow$ 0 to $I_{\max }$} 

\State Label population $P_i$ with rank and crowding distance
 
\State $\lambda(\varphi_{elite}) \leftarrow$ \textsc{Sel}(${\pi_\theta}, \mathcal{Q},P_i$)
 \hfill \footnotesize $\triangleright$ Optimization trajectory reflection
 
\State $\varphi_{c} \leftarrow$ \textsc{ConvergentSearch}(${\pi_\theta },\Phi, \lambda(\varphi_{elite})$) 
 \hfill \footnotesize $\triangleright$ Oriented tuned directive configurations  $\varphi_{c}$
 
\State $\varphi_{d} \leftarrow$ \textsc{DivergentSearch}(${\pi_\theta },\lambda(\varphi_{elite})$) 
 \hfill \footnotesize $\triangleright$ Non-oriented tuned directive configurations $\varphi_{d}$
 
\State Embed $\varphi_{c}$ and $\varphi_{d}$ into original HLS design $\lambda(\varphi)$, evaluate $\mathcal{Q}$ of $\lambda(\varphi_{c})$ and $\lambda(\varphi_{d})$

\State Adaptive population management for $P_i$
 
 \EndFor
 
 \State Select Pareto-optimal designs ${\lambda}(\varphi^*)$ through non-dominated sorting of all reported $\mathcal{Q}$

 \State \Return ${\lambda}(\varphi^*)$
\end{algorithmic}
\end{algorithm}

Current LLMs exhibit performance degradation when handling lengthy contexts and face difficulties when scaling to extensive design spaces. Furthermore, initial sampling alone inadequately captures the mapping between directive allocation strategies and their effectiveness. 
To address these limitations, we propose \textbf{QoR-Aware Adaptive Optimization} system that integrates LLM-based optimization trajectory reflection, bottleneck analysis, and design refactoring for HLS design refinement.
This system leverages convergent and divergent thinking capabilities of LLMs to expedite Pareto-optimal design acquisition while achieving broader design space coverage. Algorithm \ref{algorithm} details this process.

\textbf{Optimization Trajectory Reflection.} Following the \textit{Warm-Start} phase (lines 1-2), we construct the initial population utilizing the elite individual selection strategy resembling the NSGA-\uppercase\expandafter{\romannumeral2} framework \cite{Pareto_4}, which prioritizes superior designs while maintaining population diversity, ensuring both coverage and uniform distribution along the explored Pareto front (line 4).
For labeled designs and their QoR, we prompt the LLM to perform reflection on the optimization trajectory, thus facilitating the sensible selection of directive configurations with optimization potential.
Subsequently, a lightweight analysis examines the selected designs for data dependencies that potentially block pipeline/unroll optimization, memory access patterns to verify alignment between memory partitioning parallelism, and resource saturation to determine hardware constraint compatibility (line 5). 
This process illustrates how LLMs explore diverse optimization directions during the search process while identifying appropriate niches for trade-offs or specialized optimizations.

\textbf{Bottleneck Analysis with QoR-Aware Adaptation.} When design goals remain elusive, experts adopt a systematic approach to identify critical bottlenecks and take advantage of potential optimization opportunities rather than discard existing work altogether. 
The \textit{optimization trajectory reflection} provides valuable insights that guide subsequent LLM reasoning about promising design refinement. 
Initially, the LLM classifies designs as compute-bound or memory-bound based on QoR (line 6). For the compute-bound, the LLM progressively enhances unrolling granularity in performance-critical loops, guided by optimization trajectory analysis.
For the memory-bound, the LLM adjusts memory partition factors to meet or exceed the corresponding data access unroll factors while examining nested loops to implement coarse-grained pipelining toward outer loops.

\textbf{Divergence-Enhanced Design Refactoring.} LLMs often show hesitance to extrapolate beyond established examples and venture into unexplored design territories. To enhance exploration coverage across the entire design space, we prompt the LLM to scrutinize current hardware optimization strategies and develop novel directive configurations different from previous iterations (line 7).
Additionally, we specified the rule of first coarsely estimating the lower bound of the initiation intervals, and then applying the pipelining strategy to the innermost loop in the nested loop, while dynamically adapting memory partition factor to match loop operation characteristics. 
Through the shuffle of directive combinations under specific hardware optimization principles, LLM implements innovative optimization strategies for non-oriented refinement of the original design.

\vspace{-10pt}

\section{Evaluations \& Discussions}
\label{sec:Evaluations}
\vspace{-10pt}

In this section, we evaluate the effectiveness of iDSE in exploring Pareto-optimal designs that satisfy diverse optimization preferences. We selected 12 HLS benchmarks with varying functionality and design space dimensions $\vert \Phi \vert$ from PolyBench \cite{bench_polybench}, CHStone \cite{bench_chstone}, and MachSuite \cite{bench_machsuite}. Diverse design structures and memory footprints demonstrate the robust generalization of our approach. All synthesis was performed on Vitis HLS 2022.1 \cite{vitishls} targeting the Xilinx ZCU106 MPSoC platform. 
We compared iDSE against traditional DSE methods, including evolutionary algorithms (NSGA-\uppercase\expandafter{\romannumeral2} \cite{Pareto_4} and MOEA/D) \cite{Pareto_3}, swarm intelligence techniques (ACO) \cite{Pareto_5}, and state-of-the-art approaches including Lattice \cite{DSE_lattice} with guided local exploration and HGBO-DSE, a Bayesian optimization based on MOTPE \cite{hgbo}. iDSE consistently outperformed these methods by exploring more comprehensive and impressive Pareto fronts. Experimental results highlight the capability of iDSE as an LLM-navigated design space exploration approach that effectively unleashes the potential of LLMs in hardware optimization. The complete definitions of the benchmark design space and hyperparameter settings for heuristic-based DSE methods are detailed in Appendix \ref{app:experiment}.

\begin{table}[t]
\tiny
\centering
\caption{Comparison of iDSE effectiveness over baseline DSE methods.}
\renewcommand{\arraystretch}{1.0}
\label{table1}
\resizebox{\linewidth}{!}{ 
\begin{tabular}{@{}ccc|cccccc}
\toprule
\multicolumn{3}{c|}{HLS Design} & \multicolumn{6}{c}{ADRS of Pareto-Optimal Designs Explored by Different DSE Methods} \\ \midrule
\multicolumn{1}{c}{Benchmark} & $\vert \Phi \vert$ & \# Directives & NSGA-\uppercase\expandafter{\romannumeral2} & ACO & MOEA/D & Lattice & HGBO-DSE & \cellcolor[HTML]{E5E5FF}\textbf{iDSE(ours)}\\ \midrule
\multicolumn{1}{c}{\textit{atax}}& 4.2$\mathrm{M}$& 13& 2.3900 & 2.1073 & 0.8322 & 1.4974& 0.3070& \cellcolor[HTML]{E5E5FF}{\color[HTML]{FF0000} \textbf{0.0355}} \\
\multicolumn{1}{c}{\textit{bicg}}& 0.9$\mathrm{M}$& 12& 1.0109 & 0.2511 & 0.4540 & 4.6001& 0.2429& \cellcolor[HTML]{E5E5FF}{\color[HTML]{FF0000} \textbf{0.0497}} \\
\multicolumn{1}{c}{\textit{gemm}}& 38.5$\mathrm{M}$ & 14& 1.1061 & 0.6338 & 0.4611 & 3.9707& 0.4710& \cellcolor[HTML]{E5E5FF}{\color[HTML]{FF0000} \textbf{0.1039}} \\
\multicolumn{1}{c}{\textit{gesummv}} & 12.6$\mathrm{M}$ & 11& 0.6498 & 0.3935 & 0.3813 & 1.5178& 0.3549& \cellcolor[HTML]{E5E5FF}{\color[HTML]{FF0000} \textbf{0.0231}} \\
\multicolumn{1}{c}{\textit{mvt}} & 3.7$\mathrm{M}$& 14& 1.8678 & 1.8678 & 2.1441 & 1.9295& 0.5934& \cellcolor[HTML]{E5E5FF}{\color[HTML]{FF0000} \textbf{0.0497}} \\
\multicolumn{1}{c}{\textit{md-knn}}& 33.6$\mathrm{M}$ & 11& 0.0217 & 0.0250 & 0.0245 & 0.0115& {\color[HTML]{FF0000} \textbf{0.0068}} & \cellcolor[HTML]{E5E5FF}0.0118\\
\multicolumn{1}{c}{\textit{spmv}}& 0.3$\mathrm{M}$& 8 & 0.4744 & 0.9045 & 0.3578 & 0.1258& 0.0670& \cellcolor[HTML]{E5E5FF}{\color[HTML]{FF0000} \textbf{0.0126}} \\
\multicolumn{1}{c}{\textit{stencil2d}} & 39.0$\mathrm{K}$ & 11& 1.7985 & 0.8223 & 0.3248 & 1.0410& 0.3955& \cellcolor[HTML]{E5E5FF}{\color[HTML]{FF0000} \textbf{0.0461}} \\
\multicolumn{1}{c}{\textit{stencil3d}} & 58.7M & 21 & 1.7009  & 0.6733 & 0.4468  & 1.5011 & 0.4600 & \cellcolor[HTML]{E5E5FF}{\color[HTML]{FF0000} \textbf{0.1646}}  \\
\multicolumn{1}{c}{\textit{viterbi}} & 55.7$\mathrm{M}$ & 21& 0.1109 & 0.1743 & 0.2272 & 0.0382& 0.0316& \cellcolor[HTML]{E5E5FF}{\color[HTML]{FF0000} \textbf{0.0022}} \\
\multicolumn{1}{c}{\textit{sha}} & 12.3$\mathrm{K}$ & 8 & 0.3658 & 0.3287 & 0.2744 & 2.0663& 0.2749& \cellcolor[HTML]{E5E5FF}{\color[HTML]{FF0000} \textbf{0.0945}} \\
\multicolumn{1}{c}{\textit{autocorr}}& 27.6$\mathrm{K}$ & 10& 0.0883 & 0.0847 & 0.0621 & {\color[HTML]{FF0000} \textbf{0.0093}} & 0.0486& \cellcolor[HTML]{E5E5FF}0.0168\\ \midrule
\multicolumn{3}{c|}{\textbf{Avg Improv. over NSGA-\uppercase\expandafter{\romannumeral2}}} & \textbf{1} & \textbf{1.5376$\times$} & \textbf{2.0724$\times$} & \textbf{2.0451$\times$}& \textbf{3.7021$\times$}& \cellcolor[HTML]{E5E5FF}\textbf{25.9987$\times$} \\
\multicolumn{3}{c|}{\textbf{Geo Mean Improv. over NSGA-\uppercase\expandafter{\romannumeral2}}} & \textbf{1} & \textbf{1.3048$\times$} & \textbf{1.6839$\times$} & \textbf{1.0876$\times$}& \textbf{3.2531$\times$}& \cellcolor[HTML]{E5E5FF}\textbf{16.5955$\times$}\\ \bottomrule
\end{tabular}}
\vspace{-10pt}
\end{table}

\begin{figure*}[t]
	\centering 
\includegraphics[width=14cm]{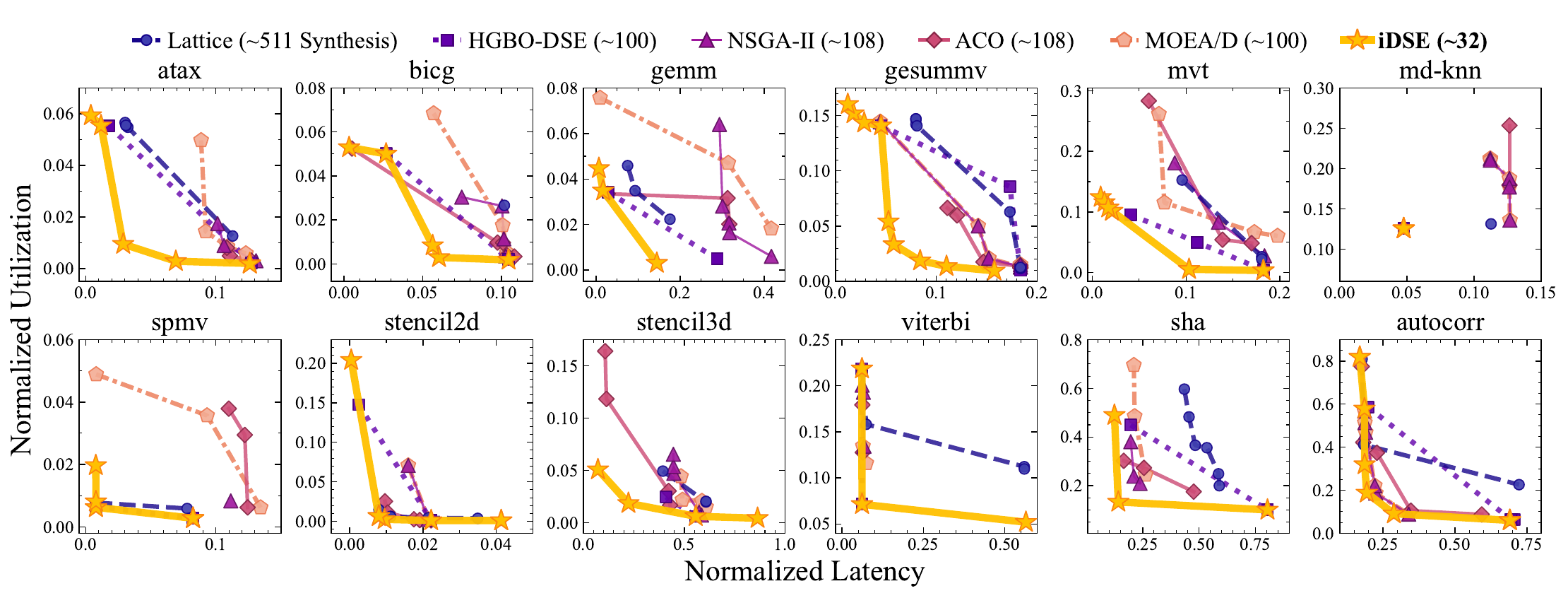} 
\vspace{-13pt}
	\caption{Comparison of explored Pareto fronts across benchmarks with different design spaces. iDSE converges to comprehensive and concave shapes with fewer search budgets (\textit{\# Synthesis}).}
	\label{experiment1}
\vspace{-18pt}
\end{figure*}

\subsection{\textls[-1]{Navigating Design Space Exploration: Elegant and Swift Approximation of Pareto Front}}

We evaluated the capability of iDSE to approximate reference Pareto fronts using ADRS (complete definition is elaborated in Appendix \ref{app:adrs}).
To ensure practicality, we extensively sampled design spaces using random sampling and specialized breadth-first search to construct the reference Pareto fronts. For smaller designs (\textit{stencil2d}, \textit{autocorr}, and \textit{sha}), we performed exhaustive exploration to construct strong reference Pareto fronts. All DSE methods explored design spaces compressed by our \textit{Feature-Driven Pruning} method to ensure fair comparison.
Table \ref{table1} demonstrates the superior performance of iDSE across most benchmarks. iDSE achieves geometric mean improvements of 16.6$\times$ over NSGA-\uppercase\expandafter{\romannumeral2}, 12.7$\times$ over ACO, 9.9$\times$ over MOEA/D, 15.3$\times$ over Lattice and 5.1$\times$ over HGBO-DSE.
While Lattice achieved optimal performance in \textit{autocorr} through local search traversal, its sensitivity to initial sampling designs and lack of global perspective limit its effectiveness in exploring trade-off curves in large memory footprints and complex design structures. Similarly, HGBO-DSE outperforms other baselines but still faces limitations when confronting vast design spaces without prior knowledge guidance. Other DSE methods compressed exploration efficiency under limited search budgets.
We present full experimental details in Appendix \ref{app:results}.

Figure \ref{experiment1} further illustrates the advantage of iDSE in multiobjective optimization, establishing more comprehensive and concave Pareto fronts with fewer search budgets (under 50 explored designs per benchmark).
Table \ref{table2} compares the number of explored designs required by different DSE methods to achieve the target ADRS. iDSE significantly enhances exploration efficiency, 
\begin{wraptable}{r}{0.435\textwidth}
\centering
\caption{Comparison of search budgets for target ADRS across different benchmarks.}
\vspace{-5pt}
\renewcommand{\arraystretch}{1.0}
\label{table2}
\scriptsize
\begin{tabular}{c|ccc>{\bfseries}c}
\toprule
DSE Methods & Poly. & Mach. & CHS. & Speedup \\ 
\midrule
NSGA-\uppercase\expandafter{\romannumeral2} & 108 & 108 & 108& 1$\times$\\
ACO& 41& 68& 83& 3.31$\times$\\
MOEA/D& 22& 61& 38& 3.56$\times$\\
Lattice & 89& 55& 72& 2.22$\times$\\
HGBO-DSE & 8& 9& 74& 14.34$\times$ \\
\rowcolor[HTML]{E5E5FF} 
\textbf{iDSE} & 4& 5& 7 & 25.07$\times$ \\ 
\bottomrule
\end{tabular}
\end{wraptable}
achieving a geometric mean speedup of 11.0$\times$ over meta-heuristic-based DSE methods and 4.4$\times$ over SOTA DSE methods.
The notable improvement in performance and efficiency of iDSE stems from the dual effectiveness of initial high-quality sampling and intelligently guided searches. LLM accelerates convergence to Pareto-optimal designs that satisfy diverse optimization preferences by prudent directive combination scheduling and parameter tuning.

\begin{table}[t]
\normalsize
\centering
\setlength{\heavyrulewidth}{1.5pt}
\caption{ADRS comparison of heuristic-based DSE methods with different initial sampling.}
\label{table3}
\renewcommand{\arraystretch}{1.2}
\resizebox{\linewidth}{!}{
\renewcommand{\arraystretch}{1.2}
\begin{tabular}{@{}c|cccc|cccc|ccc@{}}
\toprule
 & \multicolumn{4}{c|}{NSGA-\uppercase\expandafter{\romannumeral2}}& \multicolumn{4}{c|}{MOEA/D}& \multicolumn{3}{c}{ACO} \\ \cmidrule(l){2-12}
\multirow{-2.2}{*}{Benchmark} & RS & BS & LHS & \cellcolor[HTML]{E5E5FF}\textit{Warm-Start}& RS & BS & LHS& \cellcolor[HTML]{E5E5FF}\textit{Warm-Start} & RS & BS & LHS\\ \midrule
\textit{atax} & 2.3900& 1.1054& 0.5737& \cellcolor[HTML]{E5E5FF}0.3116& 0.8322& 0.4583& 0.3057& \cellcolor[HTML]{E5E5FF}0.1762& 2.1073& 0.9770& 0.7917\\
\textit{bicg} & 1.0109& 0.3685& 1.0765& \cellcolor[HTML]{E5E5FF}0.2428& 0.4540& 0.2588& 0.2573& \cellcolor[HTML]{E5E5FF}0.2758& 0.2511& 0.3938& 0.2570\\
\textit{gemm} & 1.1061& 0.5100& 1.0581& \cellcolor[HTML]{E5E5FF}0.4581& 0.4611& 0.3768& 0.3664& \cellcolor[HTML]{E5E5FF}0.2976& 0.6338& 0.4839& 0.5489\\
\textit{gesummv} & 0.6498& 0.6508& 0.9751& \cellcolor[HTML]{E5E5FF}0.5575& 0.3813& 0.4097& 0.3103& \cellcolor[HTML]{E5E5FF}0.2640& 0.3935& 0.4165& 0.3682\\
\textit{mvt} & 1.8678& 0.8544& 0.6403& \cellcolor[HTML]{E5E5FF}0.4193& 2.1441& 1.6956& 0.5464& \cellcolor[HTML]{E5E5FF}0.4976& 1.8678& 1.2207& 0.8978\\
\textit{md-knn} & 0.0217& 0.0208& 0.0064& \cellcolor[HTML]{E5E5FF}0.0064& 0.0245& 0.0184& 0.0064& \cellcolor[HTML]{E5E5FF}0.0116& 0.0250& 0.0218& 0.0064\\
\textit{spmv} & 0.4744& 1.0585& 0.8129& \cellcolor[HTML]{E5E5FF}0.0592& 0.3578& 0.8841& 0.2643& \cellcolor[HTML]{E5E5FF}0.1175& 0.9045& 0.9528& 0.5974\\
\textit{stencil2d} & 1.7985& 0.4616& 1.2631& \cellcolor[HTML]{E5E5FF}0.3622& 0.3248& 0.2991& 0.3789& \cellcolor[HTML]{E5E5FF}0.2805& 0.8223& 0.9477& 1.3129\\
\textit{stencil3d} & 1.7009& 0.6244& 1.5550& \cellcolor[HTML]{E5E5FF}0.1780& 0.4468& 0.6172& 0.2057& \cellcolor[HTML]{E5E5FF}0.2891& 0.6733& 0.6327& 0.6682\\
\textit{viterbi} & 0.1109& 0.1069& 0.0300& \cellcolor[HTML]{E5E5FF}0.0158& 0.2272& 0.1421& 0.0328& \cellcolor[HTML]{E5E5FF}0.0540& 0.1743& 0.1362& 0.1351\\
\textit{sha} & 0.3658& 0.1626& 0.7494& \cellcolor[HTML]{E5E5FF}0.1521& 0.2744& 0.0871& 0.2060& \cellcolor[HTML]{E5E5FF}0.2095& 0.0847& 0.2313& 0.2571\\
\textit{autocorr} & 0.0883& 0.0702& 0.0931& \cellcolor[HTML]{E5E5FF}0.7458& 0.0621& 0.0901& 0.0901& \cellcolor[HTML]{E5E5FF}0.0631& 0.3287& 0.0704& 0.0642
\\ \midrule
\textbf{Avg}& \textbf{1} & \textbf{1.9097$\times$} & \textbf{1.7808$\times$} & \cellcolor[HTML]{E5E5FF}\textbf{4.6109$\times$} & \textbf{2.0724$\times$} & \textbf{2.5986$\times$} & \textbf{3.7184$\times$} & \cellcolor[HTML]{E5E5FF}\textbf{4.2138$\times$} & \textbf{1.7401$\times$} & \textbf{1.6765$\times$} & \textbf{2.0451$\times$} \\
\textbf{Geo Mean} & \textbf{1} & \textbf{1.6508$\times$} & \textbf{1.3656$\times$} & \cellcolor[HTML]{E5E5FF}\textbf{3.1709$\times$} & \textbf{1.6839$\times$} & \textbf{1.9743$\times$} & \textbf{3.1445$\times$} & \cellcolor[HTML]{E5E5FF}\textbf{3.4065$\times$} & \textbf{1.3048$\times$} & \textbf{1.5094$\times$} & \textbf{1.8210$\times$} \\ \bottomrule
\end{tabular}}
\\[3pt]
\raggedright 
{\tiny 
\renewcommand{\baselinestretch}{1.2}
\noindent *~Experiments exclude ACO with \textit{Warm-Start} because swarm intelligence struggles to leverage the advantage of seed directive configurations, detailed results in Appendix \ref{ACO}. The ADRS are calculated from the average results of 5 test runs, and \textit{Warm-Start} are the average results of 5 rounds of DeepSeek-R1 invocations.\par}
\vspace{-7pt}
\end{table}

\begin{figure*}[t]
	\centering 
        \includegraphics[width=14cm]{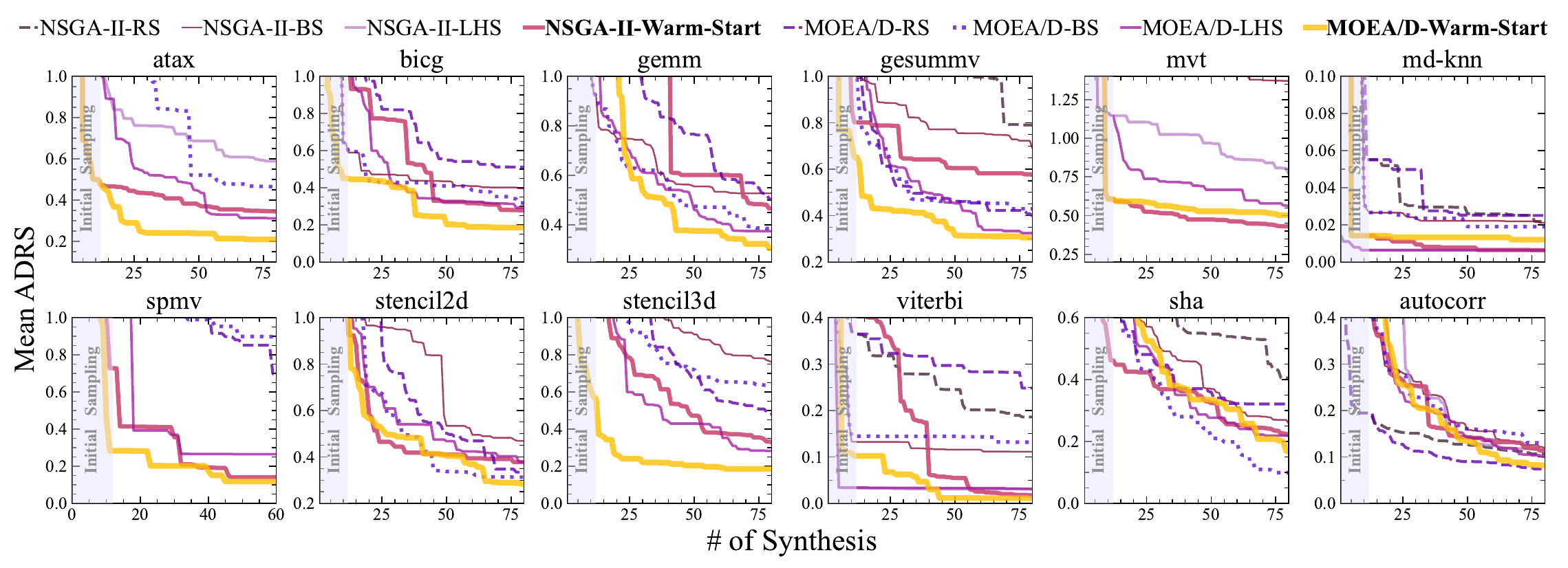} 
        \vspace{-13pt}
	\caption{Comparison among EA-based DSE methods under different initial sampling designs.}
	\label{experiment2}
\vspace{-12pt}
\end{figure*}

\subsection{DSE Warm-Start: Robust Sampling Initialization for Convergence Acceleration}

Heuristic-based DSE approaches often suffer from reduced convergence capabilities due to the lack of diverse and insightful initial sampling designs, which limits their ability to identify effective optimization directions. 
In contrast, Lattice uses U-shaped Beta sampling to strategically sample the boundaries of the design space, enabling more sensible cluster-based exploration near reference Pareto fronts. 
Therefore, we propose that initial sampling quality is a critical determinant of both overall DSE effectiveness and the convergence toward optimal designs. 
To validate our hypothesis and assess the efficacy of the \textit{Seed Directive Generation (Warm-Start)} method, we conducted comparative analyses against Random Sampling (RS), U-shaped Beta Sampling (BS), and Latin Hypercube Sampling (LHS) within heuristic-based DSE methods. As shown in Table \ref{table3}, ADRS drops substantially when seed designs more accurately approximate the reference Pareto fronts. Both BS and LHS methods consistently outperform RS across NSGA-\uppercase\expandafter{\romannumeral2}, ACO, and MOEA/D. By reasoning about HLS design structure and viable optimization space, LLM optimally constructs superior approximate Pareto fronts within limited explored designs. When paired with traditional evolutionary algorithms (EA) as search engines, our method achieves sampling effects that exceed RS by 2.5$\times$, BS by 1.8$\times$, and LHS by 1.6$\times$. For more details, please refer to Appendix \ref{app:initial}.

Figure \ref{experiment2} depicts the ADRS descent curve, showing that \textit{Warm-Start} converges more rapidly towards reference Pareto fronts, achieving lower ADRS compared to probability distribution-based alternatives. The subsequent EA-based DSE further recovers entire Pareto fronts. 
More insightfully, we observed that the improved initial sampling quality produced minimal improvements for ACO. This is probably attributed to the reliance on dynamic updates during the iteration process, which gradually overrides initial path advantages through evaporation and reinforcement processes, thereby diminishing the influence of initially superior solutions (detailed evidence in Appendix \ref{ACO}). 
In contrast, EA-based methods explicitly preserve high-quality genes, allowing the quality of seed designs to persistently guide search directions and yield more substantial improvements.

\subsection{Ablation Study: Infusing Optimization Intuition into Directive Configuration Tuning}

Traditional EAs, lacking domain expertise, struggle to capture the nuanced relationship of optimization directives applied to specific HLS designs. We enable more sensible directive configuration tuning by introducing LLM to infuse hardware optimization intuition. To evaluate our approach, we conducted ablation experiments with the following settings:

\begin{itemize}[leftmargin=*]

\item \textit{Baseline:} We prompted LLM with the HLS design and the corresponding QoR to generate directive combinations and parameters different from the initial sampling designs.

\item \textit{S1:} We incorporated information of the pruned design space to constrain exploration within reasonable boundaries, examining the effectiveness of pruning strategies on DSE.

\item \textit{S2:} We analyzed optimization trajectories of selected parents to prompt LLM to generate novel designs, examining the benefits of the \textit{optimization trajectory reflection} on the subsequent search.

\item \textit{S3:} We further integrated bottleneck analysis and design refactoring operators, examining the enhancement effects of domain-specific hardware optimization knowledge on DSE.

\end{itemize}

\begin{wrapfigure}{r}{0.45\textwidth}
    \centering
    \includegraphics[width=\linewidth]{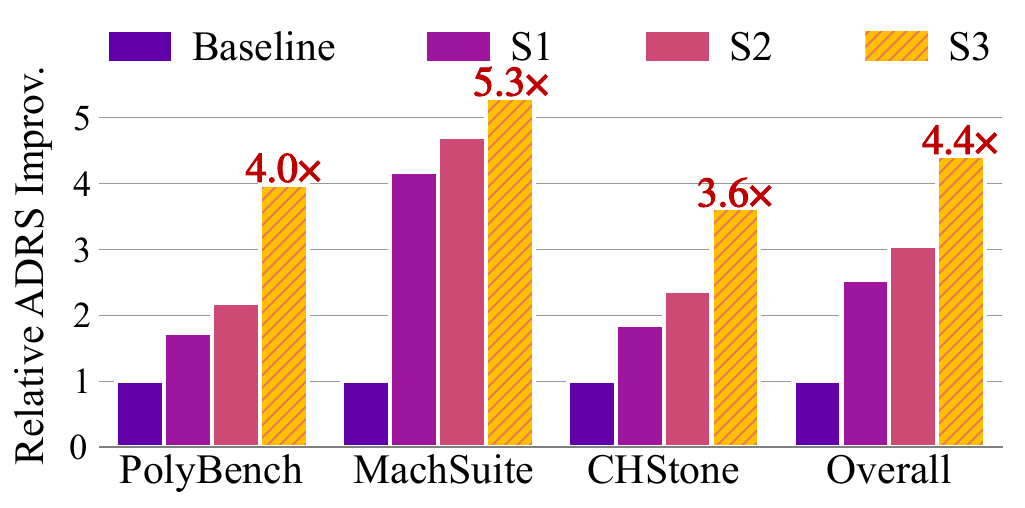}%
    \vspace{-5pt}
    \caption{Ablation of Adaptive Optimization.}
    \label{fig:ablation}
\end{wrapfigure}
The geometric mean improvement trend of ADRS in Figure \ref{fig:ablation} demonstrates that, given comparable search budgets, the proposed \textit{QoR-Aware Adaptive Optimization (S3)} system effectively leverages prior knowledge for more efficient design space exploration. 
Without proper perception of design space dimensions (\textit{Baseline}), aggressive exploration often encounters invalid designs, comprising 12.4\% of total designs, degrading the efficiency of DSE. Constraining the exploration space within the ranges established by our \textit{Feature-Driven Pruning (S1)} significantly eliminated 89.9\% invalid designs.
Furthermore, introducing \textit{optimization trajectory reflection (S2)} improved the effectiveness of optimization by 20.5\%. Injecting domain-specific knowledge into the search process \textit{(S3)} further improved the effectiveness by an additional 45.0\%. 
Additionally, our approach maintains prompt interfaces that enable further incorporation of hardware optimization preferences and resource constraints for flexible outputs.

\section{Conclusion}
\label{sec:Conclusion}
This paper presents iDSE, an effective and efficient LLM-navigated framework for automated high-level synthesis design space exploration, addressing the challenges of time-consuming hardware optimization with steep expertise barriers. 
Our approach leverages LLMs to enable elegant identification of HLS design bottlenecks and exploitation of optimization opportunities, construct representative initial sampling designs and expedite subsequent refinement within pruned design spaces. 
Experimental results substantiate that iDSE markedly outperforms traditional heuristic-based DSE methods, achieving superior approximation to the Pareto front with substantially reduced search budgets. 
Furthermore, iDSE rapidly converges to more comprehensive and compelling Pareto-optimal designs, expanding the boundary of HLS advantages in agile hardware development and optimization.

\bibliographystyle{unsrt}
\bibliography{main}

\newpage
\appendix
\label{sec:Appendix}
\startcontents

\noindent
{\fontsize{22}{30}\selectfont\bfseries Appendix\par}
\vspace{0.8em}
{\fontsize{15}{20}\selectfont\bfseries Contents\par}

\printcontents{}{0}{\setcounter{tocdepth}{2}}
\vspace*{0.8em}

\newpage

\section{iDSE Implementation}
\label{app:iDSE}

\subsection{Directive Function Description}

iDSE supports three optimization directive types in Vitis HLS \cite{vitishls}, \texttt{PIPELINE} and \texttt{UNROLL} for loops, and \texttt{ARRAY\_PARTITION} for arrays. This can be generalized to other vendor HLS tools that support customizing microarchitectures through optimization directive tuning.
Table \ref{pragma_config} summarizes these optimization directive configurations and feature vectors generated for the case in Figure \ref{workflow2}. Activating \texttt{PIPELINE} can boost throughput by overlapping time domains, while the \texttt{UNROLL} factor determines hardware loop parallelism. Meanwhile, \texttt{ARRAY\_PARTITION} splits arrays to provide sufficient data bandwidth for unrolled loops. We encode \texttt{PIPELINE} and \texttt{ARRAY\_PARTITION} types as discrete integer vectors, then combine these with unroll and partition factor vectors. Constructing feature vectors in the form of tuples by selecting specific numerical values from these directive configurations helps reduce unnecessary HLS domain-specific semantics in LLM reasoning.

\vspace{5pt}

\begin{table}[h]
\normalsize
\centering
\setlength{\heavyrulewidth}{1.5pt}
\caption{Optimization Directive Configuration and Generated Feature Vectors Example}
\label{pragma_config}
\renewcommand{\arraystretch}{1.2}
\resizebox{\linewidth}{!}{
\begin{tabular}{@{}cccc@{}}
\toprule
Structure & Optimization Directive & Config. & Examples of Feature Vector\\ \midrule
\multirow{2}{*}{Loop} & \texttt{PIPELINE} & "off", "on" & \multirow{2}{*}{\{`name': `mul', `pipeline': 1, `unroll': 2\}} \\
 & \texttt{UNROLL} & integer& \\
\midrule
\multirow{2}{*}{Array}& \multirow{2}{*} {\texttt{ARRAY\_PARTITION}} & "complete", "block", "cyclic" & \multirow{2}{*}{\{`name': `C', `type': 2, `dim': 1, `factor': 2\}} \\
 & & integer& \\ \bottomrule
\end{tabular}}
\end{table}

\vspace{5pt}

\begin{lstlisting}[caption={Vitis HLS Project Configuration for Vector Hadamard Product},label={tcl_script}]
# Project Setup
open_project vector_mul
add_files vector_mul.cpp
set_top vector_mul

# Solution Configuration
open_solution solution
set_part {xczu7ev-ffvc1156-2-e}
create_clock -period 10 -name default

# Array Partition Directives
set_directive_array_partition -type cyclic -factor 2 \
 -dim 1 "vector_mul" A
set_directive_array_partition -type cyclic -factor 2 \
 -dim 1 "vector_mul" B
set_directive_array_partition -type cyclic -factor 2 \
 -dim 1 "vector_mul" C

# Loop Pipeline Directives
set_directive_pipeline "vector_mul/mul"

# Loop Unroll Directives
set_directive_unroll -factor 2 "vector_mul/mul"

# HLS Synthesis
csynth_design
exit
\end{lstlisting}

Listing \ref{tcl_script} presents an automatically generated Tcl script within iDSE workflow, designed to optimize the vector Hadamard product HLS design shown in Figure \ref{workflow2}. Lines 12-17, 20, 23 correspond to the optimization directive insertion scripts generated from the example feature vectors in Table \ref{pragma_config}. Specifically, line 20 activates pipeline optimization for the \textit{mul} loop, while line 23 implements a factor-2 parallel unrolling for this same loop. Lines 12-17 perform array partitioning on array \textit{A, B, C}, utilizing a \texttt{Cyclic} partitioning strategy with a factor of 2 to ensure alignment with the data access throughput in the loop. Within the \texttt{ARRAY\_PARTITION} directive, \texttt{Cyclic} creates smaller arrays by interleaving elements from the original array, while \texttt{Block} creates smaller arrays from consecutive blocks of the original array.

\subsection{Specific Pruning Strategies}
\label{app:prune}

\textbf{Feature-Driven Pruning} method addresses the critical challenge of inefficient design space exploration by judiciously eliminating redundant or impractical directives, particularly when aggressive parallelism directives are configured within large design footprints.
A primary consideration involves loop structure analysis, particularly for nested loops where indivisible loop boundaries relative to optimization constraints can lead to significant performance degradation. To mitigate this, the parameter configuration is constructed based on common factor vectors derived from array and loop bounds, ensuring alignment between optimization directives and hardware-imposed constraints.

For nested loop structures, the pruner employs intelligent rule-based elimination to prevent hardware over-utilization. When the outer loop is pipelined, inner loops are automatically fully unrolled, rendering inner loop unroll factor exploration redundant. The pruner detects such scenarios and disables unnecessary unroll directives for inner loops. Additionally, loops with large trip counts trigger pruning of aggressive parallelism configurations, guided by the intuition that excessive unrolling or pipelining often results in impractical resource consumption. This strategy balances exploration efficiency with design flexibility while allowing designers to tailor pruning decisions through a prompt interface for specialized optimization preferences.

Structural constraints further refine the pruning rules. For outer loops that are not perfect (only innermost loop has content, no inter-loop logic and constant bounds), inner loop unrolling is prohibited. In multilayer nested loops, unrolling directives are automatically disabled for the outermost loop to prevent excessive hardware resource utilization. Outer loops containing multiple sub-loops within their bodies are restricted from unrolling. Furthermore, missing directives in configuration files are explicitly set to default values (e.g., unroll factor 0) to prevent undefined tool behaviors. For array partitioning, \texttt{Complete} type partitioning automatically disables partition factors.

\subsection{Screening Mechanism for Designs with Optimization Potential}

During non-dominated sorting, we calculate domination counts for each design and establish a domination relationship matrix. Subsequently, we categorize the front layers, placing non-dominated solutions into the first front, directly dominated solutions into the second front, and continuing this classification hierarchically. 
To preserve population diversity, the crowding distance calculation module evaluates solutions within identical Pareto front layers across both latency and resource consumption dimensions. After arranging the solution set along each objective dimension, crowding distances are determined by calculating Manhattan distances between adjacent solutions in the normalized objective space, with boundary designs receiving maximum priority values. 
Ultimately, a composite sorting strategy prioritizes solutions in lower front levels while preserving designs with greater crowding distances within the same level. This mechanism ensures both convergence and uniform distribution of solutions along the Pareto front.

\subsection{iDSE Hyperparameter Configurations}

iDSE employs an evolutionary algorithm driven by LLMs that employs hyperparameters shown in Table \ref{iDSE_hyperparameters}.
We specifically set the initial number of samples to 12, and the reasons for this choice are detailed in the Appendix \ref{app:sample_size}.
A fixed budget of 3 generations (\(I_{max}\)) balances design exploration effects and computational budgets.
Meanwhile, the adoption of dynamic population size (\(P_i\)) allows for gradual scaling from initial exploration (\(N_0\)) to refinement. 
A non-dominated sorting with crowding distance metrics prioritizes Pareto-optimal solutions, while controlled injection of rank-2 suboptimal candidates \(\mathcal{C}_{sub}^2\) per generation maintains solution diversity.

\begin{table}[h]
\normalsize
\centering
\setlength{\heavyrulewidth}{1.2pt}
\caption{Hyperparameter setting in iDSE.}
\label{iDSE_hyperparameters}
\renewcommand{\arraystretch}{1.2}
\resizebox{\linewidth}{!}{
\begin{tabular}{@{}c|c|c@{}}
\toprule
Hyperparameter & Description& Value \\ 
\midrule
\(I_{max}\) & Number of evolutionary iterations& 3\\
\(N_0\) & Initial sampling designs for DSE warm-start & 12\\
\(P_i\) & Population size per iteration stage: {[}Init, Iteration, ...{]} & {[}12,$\sim$12{]}\\
\({C}_{sub}^2\)& Suboptimal candidates selected from Rank 2 in Pareto front & 3\\
\bottomrule
\end{tabular}}
\end{table}

\section{Prompt Engineering}
\label{app:Prompt}

This section presents our comprehensive approach to prompt LLMs in \textbf{iDSE} workflow. 
The following subsections detail our prompt architecture across six key LLM agents of the iDSE workflow. Each component represents a key stage in our automated DSE framework, transforming the traditional DSE methods to a more flexible LLM-navigated approach that ensures workflow robustness while remaining automated.
We emphasize that the prompts we designed are not tailored to any specific benchmark. A unified prompt template is applied across all benchmarks to ensure the generalization of our methodology across diverse HLS designs. This approach objectively demonstrates the inherent capacity of LLMs in hardware optimization and design space exploration.

\subsection{Prompt for Feature Extractor}

\begin{figure*}[h]
	\centering 
 \includegraphics[width=14cm]{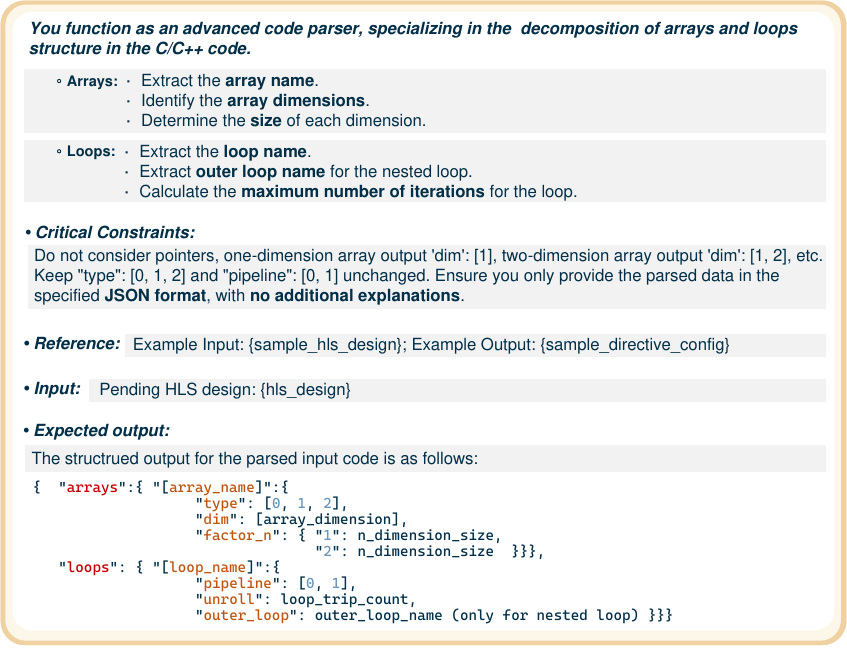} 
 \vspace{-15pt}
	\caption{Prompt for Feature Extractor.}
	\label{prompt1}
\end{figure*}

In Figure \ref{prompt1}, we present a few-shot example along with task descriptions and a structured output format used to prompt the LLM to extract key HLS design structure features. This step replaces the traditional preprocessing of the DSE method, which typically relies on the LLVM compiler to derive necessary structure information, thereby simplifying system deployment. Subsequently, a script identifies all common divisors of the target arrays and loop boundaries, thereby enabling the construction of the configurations for the optimization directives.

\vspace{5pt}

\subsection{Prompt for Feature-Driven Pruning}

\begin{figure*}[t]
	\centering 
 \includegraphics[width=14cm]{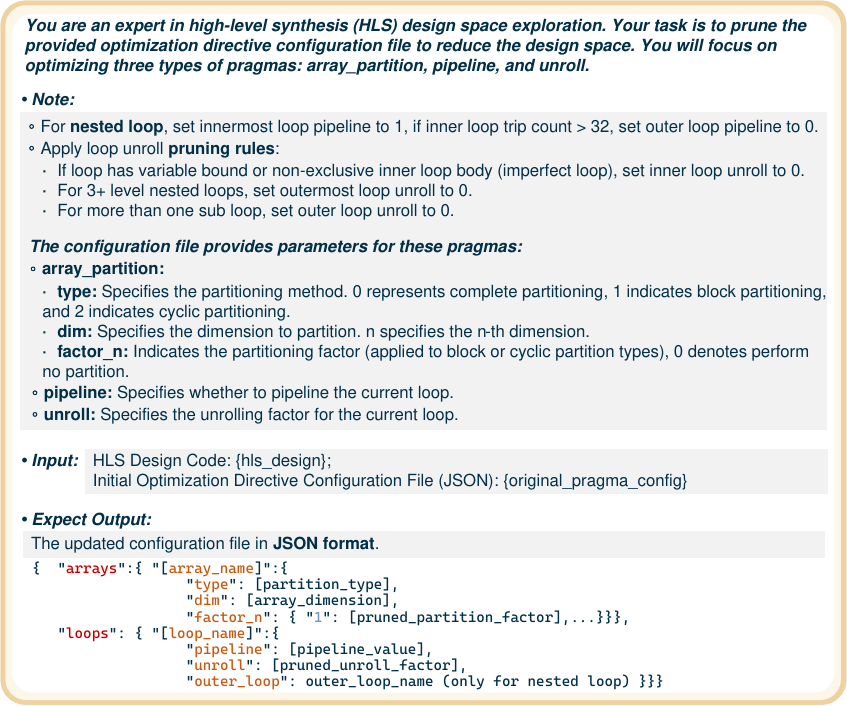} 
 \vspace{-15pt}
	\caption{Prompt for Feature-Driven Pruning.}
	\label{prompt2}
\end{figure*}

In Figure \ref{prompt2}, we prompt the LLM to leverage auxiliary pruning information in the structured JSON output from the previous step, thereby removing invalid (excessive parallelism leads to synthesis failure or duplication of optimization effects) configurations for certain optimization directives. Furthermore, we employ additional scripts to detect and eliminate potentially invalid designs within the pruned design space, as the LLM alone may not fully capture the complex interdependencies among directives. This hybrid pruning strategy preserves full automation while ensuring operational robustness throughout the optimization workflow.

\vspace{5pt}

\subsection{Prompt for Seed Directive Generation}

Figure \ref{prompt3} details our hierarchical prompt architecture integrating task descriptions, sampling objectives, and functional descriptions of optimization directives. The sampling objectives are organized through three meticulously defined optimization regimes:

\begin{itemize}[leftmargin=*]

\item \textbf{Prioritize Performance.} Maximize computational throughput through aggressive parallelism, accepting elevated resource utilization.

\item \textbf{Prioritize Resource Utilization.} Minimize hardware consumption via conservative parallelism allocation, prioritizing area efficiency over high performance.

\item \textbf{Performance-Cost Trade-off.} Identify unroll factor configurations that strike a feasible balance between performance and hardware resource budgets.

\end{itemize}

These distinct objectives help ensure that the initial sampling designs achieve broad coverage of the Pareto front. 
Furthermore, the embedded directive specifications ground LLM reasoning in domain-specific HLS constraints, and bridge the semantic gap between natural language instructions and hardware synthesis requirements through structured knowledge injection.

\begin{nolinenumbers}
\vfill
\noindent\adjustbox{trim=0 {0.62\height} 0 0, clip}{\includegraphics[width=\textwidth]{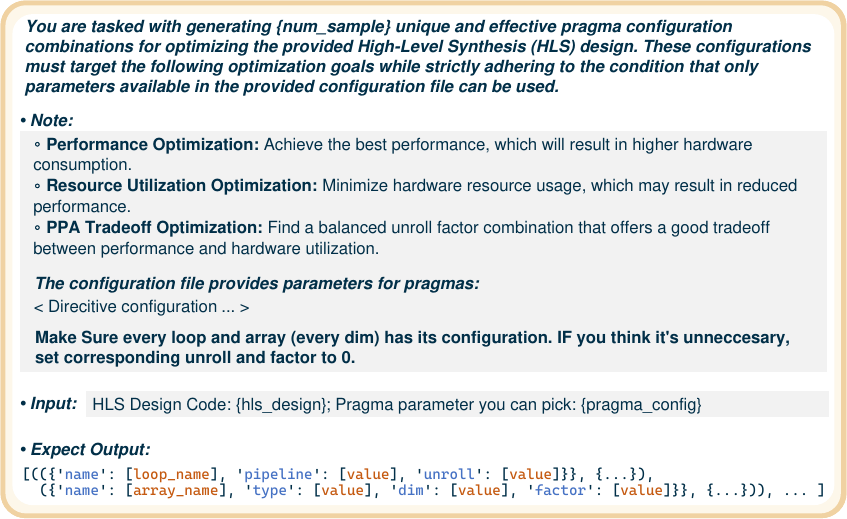}}

\begin{figure}[t]
\adjustbox{trim=0 0 0 {0.38\height}, clip}{
\includegraphics[width=\textwidth]{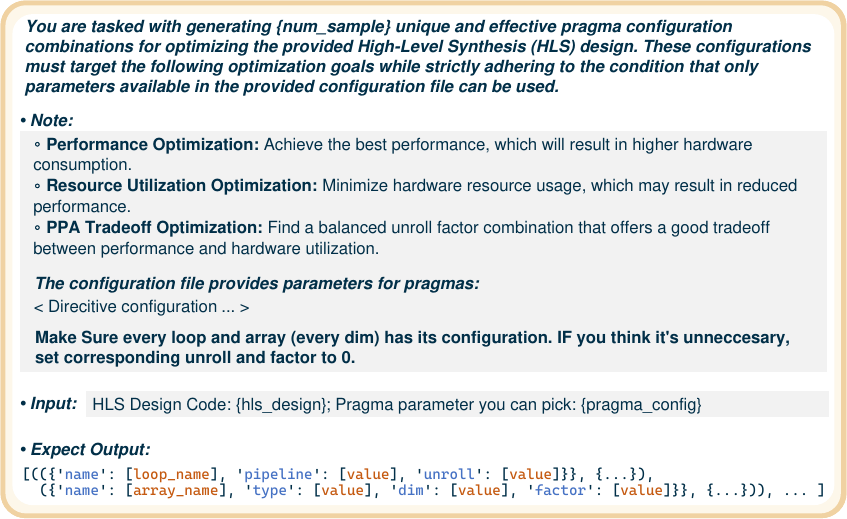}
}
\vspace{-15pt}
\caption{Prompt for Seed Directive Generation.} 
\label{prompt3}
\end{figure}
\newpage
\end{nolinenumbers}

\subsection{Prompt for QoR-Aware Adaptive Optimization}

\begin{figure*}[h]
	\centering 
 \includegraphics[width=14cm]{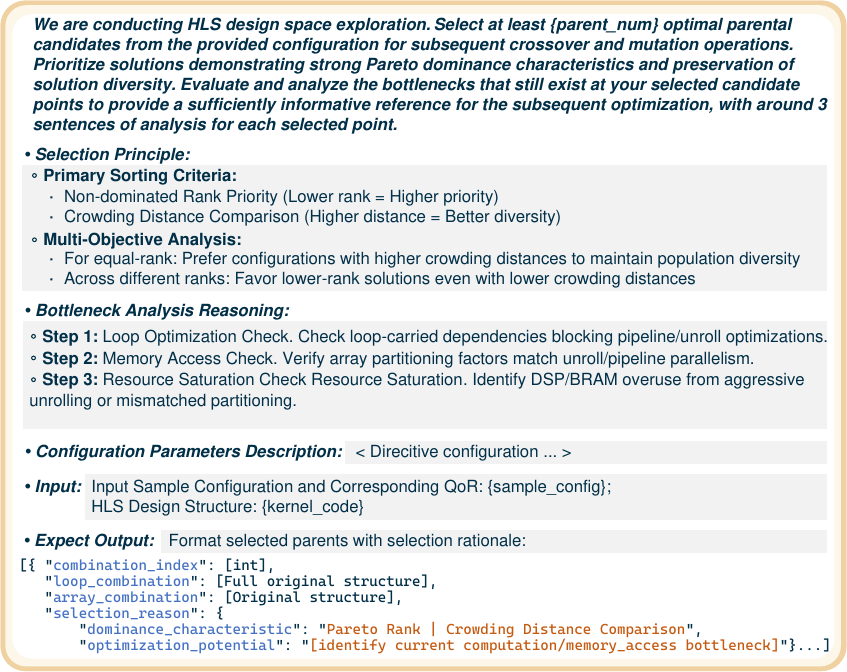} 
 \vspace{-15pt}
	\caption{Prompt for optimization trajectory reflection.}
	\label{prompt4}
 \vspace{-10pt}
\end{figure*}

In this section, we present in detail the prompts in our \textit{QoR-Aware Adaptive Optimization} system.
In Figure \ref{prompt4}, the hierarchical structure of prompt for \textbf{optimization trajectory reflection} comprises:

\begin{itemize}[leftmargin=*]

\item \textbf{Parent Selection Mechanism.} A rule-based prioritization balancing Pareto dominance and solution diversity for parental candidate selection.

\item \textbf{Bottleneck Identification.} Systematic bottleneck identification targeting loop optimization, memory access patterns, and resource saturation thresholds to guide iterative refinements.

\item \textbf{QoR-Aware Adaptation}: Synthesized circuit performance metrics (e.g., latency and resource utilization) link directive configurations to hardware implementation evaluations, allowing LLMs to trace optimization causality quantitatively.

\end{itemize}

\begin{figure*}[h]
	\centering 
 \includegraphics[width=14cm]{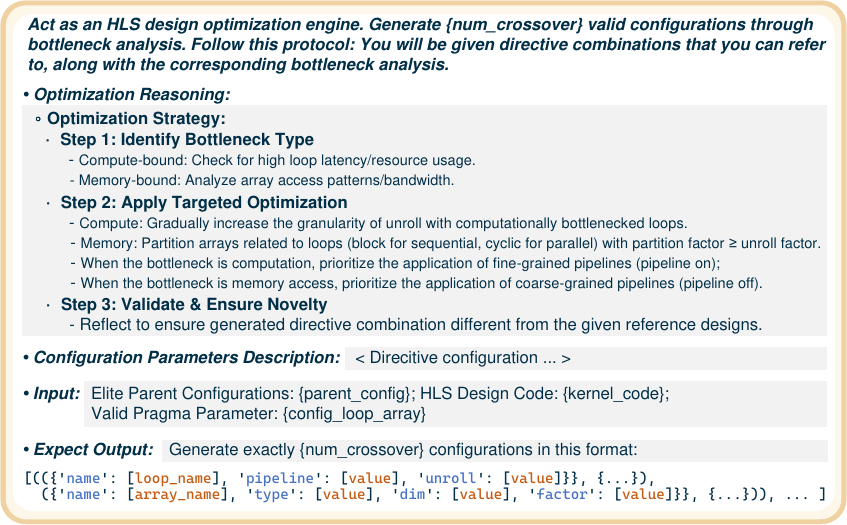} 
 \vspace{-15pt}
	\caption{Prompt for bottleneck analysis with QoR-aware adaptation.}
	\label{prompt5}

\end{figure*}

In Figure \ref{prompt5}, we prompt the LLM for \textbf{bottleneck analysis with QoR-aware adaptation}. 
It first identify compute/memory bottlenecks in selected HLS designs with optimization potential, and then perform oriented optimizations. To maintain genetic stability, we enforce a parameter preservation by retaining a sufficient subset of parent configurations. 

\vspace{10pt}

\vspace{10pt}

\begin{figure*}[h]
	\centering 
 \includegraphics[width=14cm]{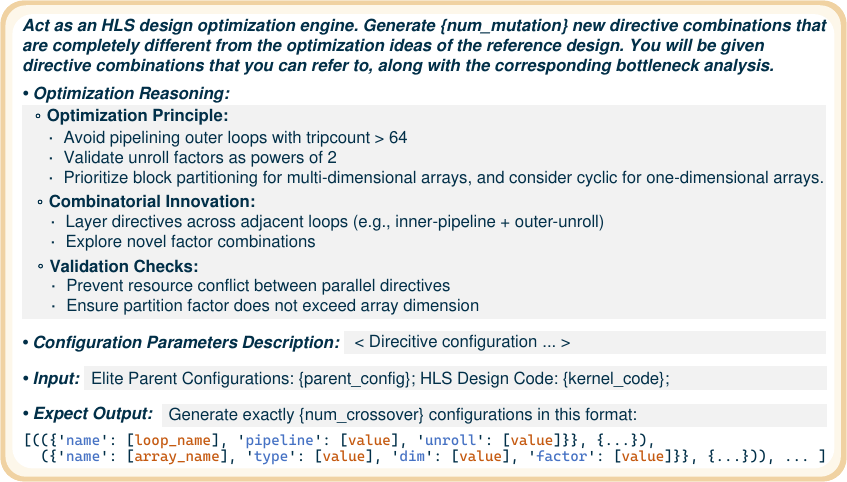} 
 \vspace{-15pt}
	\caption{Prompt for divergence-enhanced design refactoring.}
	\label{prompt6}
 \vspace{-10pt}
\end{figure*}

In Figure \ref{prompt6}, we prompt the LLM for \textbf{divergence-enhanced design refactoring} to move beyond the constrained design space and generate novel directive configurations. This approach generates configurations distinct from selected parent patterns. We prioritize loop optimizations while enforcing hardware scheduling constraints to ensure design feasibility. 
For example, restrictions are imposed on outer loop operations through trip count threshold control, enforcing the power-of-two property of the unfolding factor, and differentiating partition type selection based on the array dimension property.

\vspace{10pt}

\section{Experiment Details}
\label{app:experiment}

All experiments were conducted on an Intel Xeon Platinum 8378A server running Ubuntu 20.04.6. The designs were implemented using Vitis HLS 2022.1, targeting the Xilinx ZCU106 MPSoC platform under a maximum synthesis time constraint of 20 minutes. Evaluation of explored directive configurations exceeding this threshold typically resulted from excessively aggressive parallelization strategies that surpassed the available hardware resources.

\subsection{Evaluation Metrics}
\label{app:adrs}

The average distance to reference set (ADRS) is a widely adopted metric in High-Level Synthesis (HLS) to evaluate the quality of design space exploration (DSE). This metric quantifies the gap between an explored Pareto front \(P_E\), generated by a DSE method, and a reference Pareto front \(P_R\), which represents the best-known optimal solutions. ADRS is defined as the average normalized distance from each design in \(P_R\) to its closest counterpart in \(P_E\). Formally, the metric is computed by averaging the minimum relative degradation across all objectives for every reference design, ensuring a reasonable comparison that accounts for the latency (\(Lat\)) and resource utilization (\(Util\)) trade-offs.

\begin{equation}\label{eq3} 
ADRS({P}_E, {P}_R) = \frac{1}{|{P}_R|} \sum_{\lambda(\varphi^\gamma) \in {P}_R} \min_{\lambda(\varphi^\omega) \in {P}_E} d
\end{equation}

The distance \(d(\cdot)\) in ADRS is designed to measure the relative degradation of solutions in \(P_E\) compared to those in \(P_R\). For a reference design \(\lambda(\varphi^\gamma) \in P_R\), the distance to its nearest neighbor \(\lambda(\varphi^\omega) \in P_E\) is calculated by taking the maximum relative difference in latency and resource utilization. This difference is normalized by the objective values of reference design to ensure scale invariance, avoiding bias toward objectives with large ranges. Critically, \(d(\cdot)\) only penalizes solutions in \(P_E\) that underperform relative to \(P_R\). If a design in \(P_E\) dominates or matches the reference design in both objectives, the distance is zero. This evaluation aligns with practical DSE goals, where the primary focus is to approximate or exceed the reference front rather than explore regions beyond it.

\begin{equation}\label{eq3} 
d = \max \left\{ 0, 
\frac{Lat (\mathcal{H}, \lambda(\varphi^\omega)) - 
Lat (\mathcal{H}, \lambda(\varphi^\gamma))} 
{Lat(\mathcal{H}, \lambda(\varphi^\gamma))}, 
\frac{Util (\mathcal{H}, \lambda(\varphi^\omega)) - 
Util (\mathcal{H}, \lambda(\varphi^\gamma))}
{Util (\mathcal{H}, {\lambda(\varphi^\gamma))}} \right\}
\end{equation}

ADRS is suitable for HLS DSE due to the inherent irregularities in Pareto fronts generated during exploration. In HLS, the interplay between compiler optimizations, resource constraints, and latency-performance trade-offs often leads to non-smooth, discontinuous, or clustered Pareto fronts. Traditional metrics like hypervolume (HV), which measures the volume of the objective space dominated by a Pareto front, struggle to provide reliable comparisons in such scenarios. 
The sensitivity of HV to the shape and continuity of the front requires convexity and smoothness for meaningful interpretation, rendering it less effective when fronts exhibit abrupt transitions or sparse distributions. In contrast, ADRS circumvents these limitations by focusing on pairwise proximity between \(P_E\) and \(P_R\), making it robust to irregularities in the front geometry.

In the context of HLS design evaluation, the latency (\(Lat\)) is derived from synthesis timing analysis rather than post-implementation routing delays. This metric captures intrinsic circuit performance determined by logic-level optimization decisions, eliminating variations introduced by layout and wiring during the implementation phase. For resource utilization (\(Util\)), we define it as a weighted sum of critical FPGA schedulable resources. The utilization metric is formulated as:

\begin{equation}\label{eq_util}
Util (\mathcal{H}, \lambda(\varphi)) = W_{LUT} \cdot LUT + W_{FF} \cdot FF + W_{DSP} \cdot DSP + W_{BRAM} \cdot BRAM
\end{equation}

where $LUT$, $FF$, $DSP$, and $BRAM$ represent the normalized usage ratios of lookup tables, flip-flops, digital signal processors, and block RAMs, respectively. The weighting coefficients (\(W_{LUT}=0.3\), \(W_{FF}=0.25\), \(W_{DSP}=0.3\), \(W_{BRAM}=0.05\)) prioritize DSP and LUT resources due to their stronger correlation with computational throughput, while BRAM allocation is discounted as memory blocks are typically optimized separately through dedicated directives. 
By calibrating weights to specific hardware resources, it formalizes how HLS compiler decisions inherently balance computational density and memory bandwidth availability during design space exploration.

\vspace{20pt}

\subsection{Benchmark Descriptions and Design Structures}

\vspace{5pt}

\begin{table}[h]
\normalsize
\centering
\caption{Benchmark description with design structure.}
\label{Benchamark}
\renewcommand{\arraystretch}{1.2}
\resizebox{\linewidth}{!}{
\renewcommand{\arraystretch}{1.2}
\begin{tabular}{@{}c|ccc@{}}
\toprule
Benchmark & Description& \# Loop & \# Array \\ \midrule
\textit{atax} & Dual matrix-vector multiplications $y = A^T(Ax)$ & 4& 4 \\
\textit{bicg} & Biconjugate gradient stabilized method $q = Ap$, $s = A^Tr$ & 3& 5 \\
\textit{gemm} & BLAS general matrix multiply $C_{out} = \alpha AB + \beta C$ & 4& 3 \\
\textit{gesummv} & Summed matrix-vector multiplications $y = \alpha Ax + \beta Bx$ & 2& 5 \\
\textit{mvt} & Dual matrix-vector products $x1 = x1 + Ay1$, $x2 = x2 + A^Ty2$ & 4& 5 \\
\textit{md-knn} & Molecular dynamics force computation with neighbor lists & 2& 7 \\
\textit{spmv} & Sparse matrix-vector multiplication& 2& 4 \\
\textit{stencil2d} & 2D convolution with 3x3 kernel& 4& 3\\
\textit{stencil3d} & 3D stencil computation with two coefficients& 9& 3 \\
\textit{viterbi} & Dynamic programming for optimal hidden state sequence in HMM & 7& 6 \\
\textit{sha} & SHA-1 cryptographic hash computation rounds & 6& 3 \\
\textit{autocorr} & Autocorrelation computation with scaling and lag products& 6& 0 \\ \bottomrule
\end{tabular}}
\end{table}

\subsection{Heuristic-Based DSE Configurations}

\vspace{5pt}

\begin{table}[h]
\footnotesize
\centering
\caption{Baseline DSE method hyperparameter setting.}
\label{baseline}
\renewcommand{\arraystretch}{1.2}
\begin{tabular}{@{}c|c@{}}
\toprule
Baseline & Hyperparameter Settings\\ \midrule
NSGA-\uppercase\expandafter{\romannumeral2}& \begin{tabular}[c]{@{}c@{}} \textit{Initial samples = 12, Total generations = 8, } \\ \textit{Crossover probability = 0.9, Mutation probability = 0.3} \end{tabular}
\\ \midrule
MOEA/D& \begin{tabular}[c]{@{}c@{}} \textit{Population size=12, Max evaluations=100, Neighbor size=5,} \\ \textit{Mutation rate=0.1, Tchebycheff decomposition}\end{tabular}
\\ \midrule
ACO& \begin{tabular}[c]{@{}c@{}} \textit{Iterations=8, Ants=12, Evaporation rate=0.1, } \\ \textit{Pheromone update Q=100, Initial pheromone=1.0} \end{tabular}
\\ \midrule
Lattice \cite{DSE_lattice} & \begin{tabular}[c]{@{}c@{}} \textit{Initial Beta sampling ($\alpha$=0.1, $\beta$=0.1), Max samples=100,} \\ \textit{Lattice radius=0.5,} \textit{SphereTree neighbor search}\end{tabular}
\\ \midrule
HGBO \cite{hgbo} & \begin{tabular}[c]{@{}c@{}} \textit{LHS init trials=10, Total trials=100,} \\ \textit{EHVI candidates=24, Prior weight=1.0} \end{tabular}
\\ \bottomrule
\end{tabular}
\end{table}

\section{Supplementary Results}
\label{app:results}

\subsection{Determination of Initial Sample Sizes}
\label{app:sample_size}

\begin{figure*}[h]
	\centering 
 \includegraphics[width=14cm]{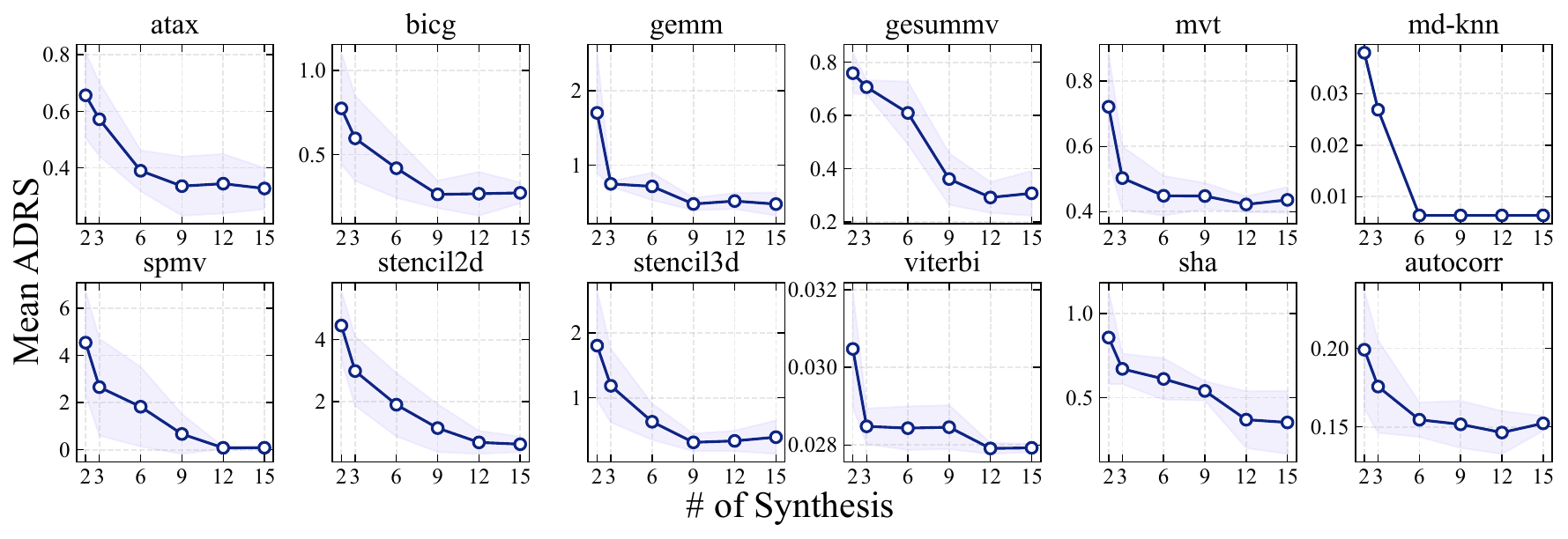}
	\caption{Optimal initial sample size selection for efficient Pareto front coverage}
	\label{sampling_size}
\end{figure*}

Our experimental observations reveal that increased initial sampling quantities generally enhance Pareto front coverage across benchmarks, as indicated by the progressive convergence toward reference fronts through diminishing ADRS values, as shown in Figure \ref{sampling_size}. While this trend demonstrates that approximation quality improves with sample size, strictly monotonic ADRS reduction does not universally occur. Furthermore, excessive sampling leads to output truncation and quality degradation due to the inherent limitations in the long-context processing capabilities of LLM. 
Through evaluation of the trade-off between coverage and computational complexity, we identified an optimal configuration of \textbf{12} initial samples that effectively balances approximation fidelity with computational efficiency. To ensure fair comparison across experimental evaluations, we uniformly applied this sample quantity across various heuristic-based DSE methods. This initialization strategy facilitates effective warm-start exploration while maintaining minimal computational budgets, highlighting both the efficacy of our sampling approach and the overall efficiency of \textbf{iDSE}.

\subsection{Construction of Pareto Front through Initial Single-Batch Sampling}
\label{app:initial}

In this section, we present detailed analysis of the data shown in Table \ref{table2}. For our experiments, we configured NSGA-\uppercase\expandafter{\romannumeral2} with 12 initial sampling designs for each benchmark, eventually reaching approximately 108 populations. 
The ADRS explored during the DSE process serves as our comparative baseline. Table \ref{table88} enumerates the number of explored designs required by ACO, MOEA/D, Lattice, and HGBO-DSE to reach or fall below the target ADRS. 
When these DSE methods failed to meet the target ADRS within the limited search budget, we defined the result as the maximum population size of the baseline method.
The results demonstrate that iDSE significantly outperforms other DSE methods, achieving efficiency improvements of 25.1$\times$ over NSGA-\uppercase\expandafter{\romannumeral2}, 7.6$\times$ over ACO, 7.0$\times$ over MOEA/D, 11.3$\times$ over Lattice, and 1.7$\times$ over HGBO-DSE. 
Notably, we observed that the ADRS threshold established in the baseline was consistently achieved across all benchmarks during just the \textit{Warm-Start} phase of iDSE. 
This finding primarily reveals that our LLM-guided \textbf{Seed Directive Generation} method can construct superior Pareto fronts through single-batch sampling. 
Furthermore, it highlights the great potential of iDSE for enhancing DSE efficiency, surpassing subsequent \textit{adaptive optimization} phases by providing designers with insightful designs in earlier stages.

\vspace{10pt}

\begin{table}[h]
\normalsize
\centering
\setlength{\heavyrulewidth}{1.5pt}
\caption{Number of explored designs required by different DSE methods to reach target ADRS.} 
\label{table88} 
\renewcommand{\arraystretch}{1.2}
\setlength{\tabcolsep}{3.5pt}
\resizebox{\linewidth}{!}{ 
\begin{tabular}{@{}c|ccccc|ccccc|cc|cc@{}}
\toprule
& \multicolumn{5}{c|}{PolyBench \cite{bench_polybench}}& \multicolumn{5}{c|}{MachSuite \cite{bench_machsuite}} & \multicolumn{2}{c|}{CHStone \cite{bench_chstone}} & \multicolumn{2}{c}{\textbf{Speedup}}\\ \cmidrule(l){2-15} 
\multirow{-2}{*}{DSE Methods} & \textit{atax} & \textit{bicg} & \textit{gemm} & \textit{gesummv} & \textit{mvt} & \textit{md-knn} & \textit{spmv} & \textit{stencil2d} & \textit{stencil3d} & \textit{viterbi} & \textit{sha} & \textit{autocorr} & \textbf{Avg}& \textbf{Geo Mean} \\ \midrule
NSGA-\uppercase\expandafter{\romannumeral2} & 108 & 108 & 108 & 108& 108& 108 & 108 & 108& 108& 108& 108& 108 & \textbf{1}& \textbf{1}\\
ACO & 19& 2 & 9 & 67 & 108& 108 & 108 & 9& 8& 108& 61 & 105 & \textbf{8.80$\times$} & \textbf{3.31$\times$} \\
MOEA/D& 18& 14& 9 & 55 & 13 & 108 & 46& 20 & 25 & 108& 39 & 36& \textbf{4.65$\times$} & \textbf{3.56$\times$} \\
Lattice & 13& 108 & 108 & 108& 108& 93& 2 & 59 & 99 & 23 & 108& 36& \textbf{6.59$\times$} & \textbf{2.22$\times$}\\
HGBO-DSE& 3 & 2 & 4 & 25 & 5& 1 & 1 & 11 & 21 & 9& 81 & 66& \textbf{32.40$\times$} & \textbf{14.34$\times$}\\
\rowcolor[HTML]{E5E5FF} 
iDSE& 3 & 3 & 6 & 4& 5& 5 & 1 & 6& 3& 10 & 6& 7 & \textbf{30.54$\times$} & \textbf{25.07$\times$} \\ \bottomrule
\end{tabular}}
\end{table}

\vspace{20pt}

\begin{figure*}[h]
	\centering 
 \includegraphics[width=14cm]{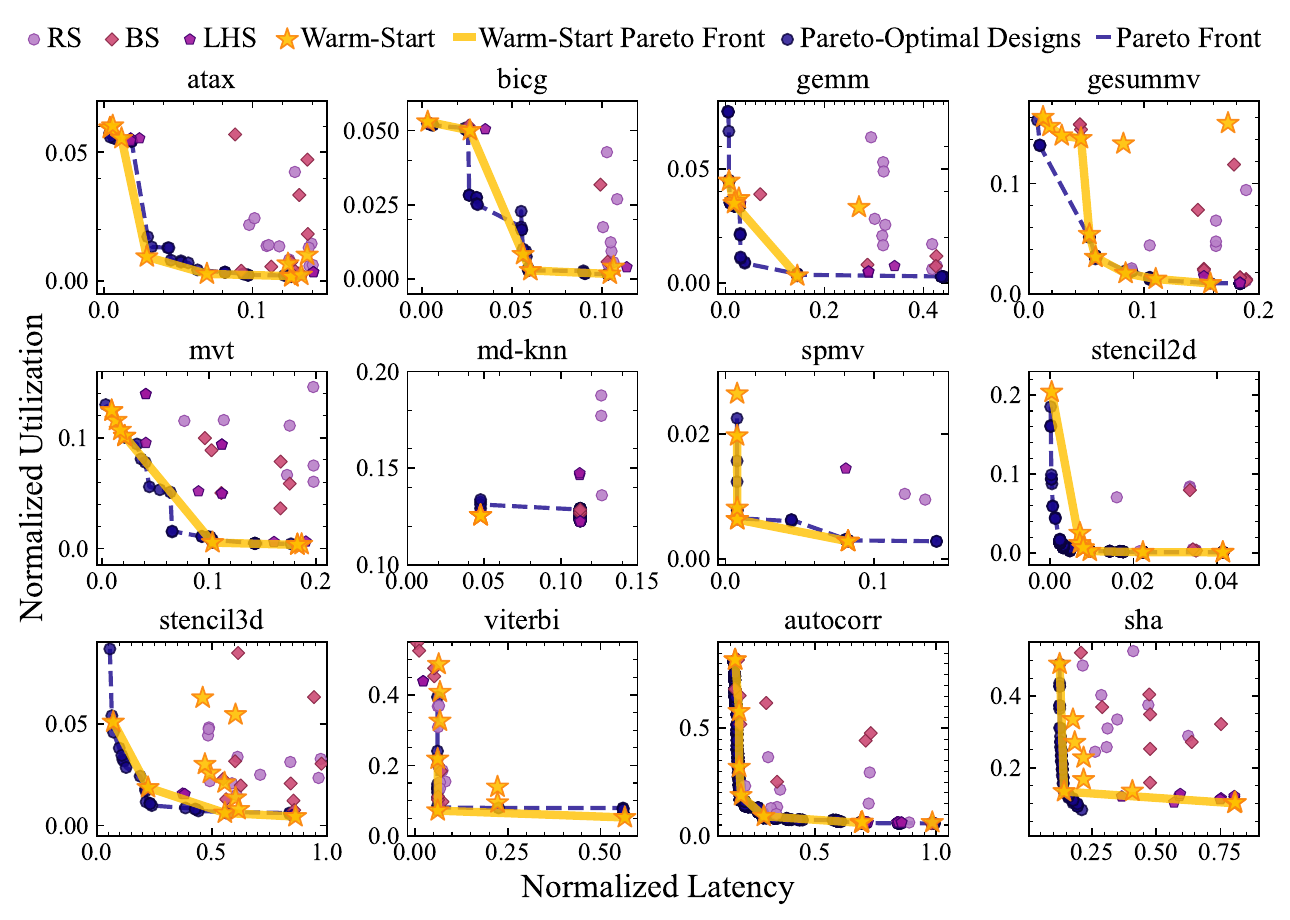} 
 \vspace{-15pt}
	\caption{Comparison of Pareto fronts constructed by different initial sampling designs}
	\label{sampling_front}
\end{figure*}

Figure \ref{sampling_front} illustrates the initial explored Pareto fronts obtained from different benchmarks under the previously determined number of initial samples. As shown, the \textit{Random Sampling (RS)}, \textit{U-shaped Beta Sampling (BS)}, and \textit{Latin Hypercube Sampling (LHS)} methods depend solely on probabilistic distributions, producing sparsely distributed and irregularly scattered designs. Consequently, these methods struggle to achieve sufficiently broad coverage or to closely approximate the reference Pareto front. 
In contrast, our \textit{Seed Directive Generation} approach leverages domain-specific prior knowledge to guide single-batch sampling, 
shaping a more comprehensive and more concave Parent front that incorporates multiple promising design configurations. 
This comparative analysis confirms that, even during the preliminary stages of exploration, our approach achieves close alignment with the ground-truth Pareto front while incurring minimal HLS compilation overhead. The results underscore the critical role of representative seed designs in accelerating early-phase design space exploration without compromising solution quality. iDSE significantly improves exploration efficiency through initial sampling warm-start for rapid approximation of the reference Pareto fronts.
This indicates that iDSE can rapidly converge, thereby reducing the time overhead required for hardware optimization while avoiding potential degradation issues that might arise from long-context prompts to LLMs.

\vspace{5pt}

\subsection{Impact of Initial Sampling and Subsequent Search on Heuristic-Based DSE}
\label{ACO}

To further validate our hypothesis that retaining advantageous genetic factors (partial parallelism for arrays, loops) can effectively leverage evolutionary algorithms for DSE in hardware optimization tasks, we decomposed the data in Table \ref{table3} into two distinct phases: an initial sampling phase (Table \ref{table71}) and a subsequent search phase (Table \ref{table81}). Our analysis reveals that the quality of sampling designs improves proportionally with the enhanced rationality of the sampling methodology. Notably, NSGA-\uppercase\expandafter{\romannumeral2} demonstrates a dependency between the effectiveness of its subsequent search and the quality of initial sampling, suggesting that this heuristic successfully preserves superior genetic factors during optimization to drive efficient DSE.
In contrast, in ACO, the ADRS values obtained during the search phase exhibit no clear correlation with the initial sampling quality. 
The improvements in the quality and dispersion of the initial sampling design provided by BS and LHS compared to RS are not reflected in the subsequent search phases of ACO, 
\begin{wraptable}{r}{0.5\textwidth}
\scriptsize
\centering
\setlength{\heavyrulewidth}{1.2pt}
\vspace{5pt}
\caption{ADRS comparison of different initial sampling methods.}
\vspace{-5pt}
\label{table71}
\renewcommand{\arraystretch}{1.2}
\begin{tabular}{@{}c|ccc
>{\columncolor[HTML]{E5E5FF}}c}
\toprule
\multicolumn{1}{l|}{Benchmark} & RS & BS & LHS & \textit{Warm-Start} \\ \midrule
\textit{atax} & 2.3900& 1.1054& 0.5737& 0.3116\\
\textit{bicg} & 1.0109& 0.3685& 1.0765& 0.2428\\
\textit{gemm} & 1.1061& 0.5100& 1.0581& 0.4581\\
\textit{gesummv}& 0.6498& 0.6508& 0.9751& 0.5575\\
\textit{mvt}& 1.8678& 0.8544& 0.6403& 0.4193\\
\textit{md-knn} & 0.0217& 0.0208& 0.0064& 0.0064\\
\textit{spmv} & 0.4744& 1.0585& 0.8129& 0.0592\\
\textit{stencil2d}& 1.7985& 0.4616& 1.2631& 0.3622\\
\textit{stencil3d}& 1.7009& 0.6244& 1.5550& 0.1780\\
\textit{viterbi}& 0.1109& 0.1069& 0.0300& 0.0158\\
\textit{sha}& 0.3658& 0.1626& 0.7494& 0.1521\\
\textit{autocorr} & 0.0883& 0.0702& 0.0931& 0.7458\\ \midrule
\textbf{Avg}& \textbf{1} & \textbf{1.9097$\times$} & \textbf{1.7808$\times$} & \textbf{4.6109$\times$} \\
\textbf{Geo Mean} & \textbf{1} & \textbf{1.6508$\times$} & \textbf{1.3656$\times$} & \textbf{3.1709$\times$} \\ \bottomrule
\end{tabular}
\end{wraptable}
as evidenced by the lack of significant differences among the indicators in the last three columns of Table \ref{table81}. Unlike evolutionary algorithms that directly inherit and recombine population genes through crossover and mutation operations, ACO relies on the accumulation and evaporation of pheromone trails. While initial sampling designs may provide some prior information, the algorithm continuously weakens the influence of historical paths through pheromone evaporation mechanisms, while dynamically covering trajectories in low-quality regions through path reconstruction. This mechanism causes ACO to rely more on real-time feedback during the iteration process rather than the static distribution of the initial population.
This empirical observation reinforces the rationale behind our methodology, which combines LLM-enhanced initial sampling with subsequent genetic-inspired search algorithms to effectively explore high-quality designs.

\begin{table}[t]
\normalsize
\centering
\setlength{\heavyrulewidth}{1.5pt}
\caption{ADRS comparison of search efficiency.}
\label{table81}
\renewcommand{\arraystretch}{1.2}
\resizebox{\linewidth}{!}{ 
\begin{tabular}{@{}c|cccc|cccc|ccc@{}}
\toprule
 & \multicolumn{4}{c|}{NSGA-\uppercase\expandafter{\romannumeral2}} & \multicolumn{4}{c|}{MOEA/D} & \multicolumn{3}{c}{ACO} \\ \cmidrule(l){2-12} 
\multirow{-2}{*}{Benchmark} & RS & BS & LHS & \cellcolor[HTML]{E5E5FF}\textit{Warm-Start}& RS & BS & LHS & \cellcolor[HTML]{E5E5FF}\textit{Warm-Start}& RS & BS & LHS \\ \midrule
\textit{atax}& 2.3900& 1.1054& 0.5767& \cellcolor[HTML]{E5E5FF}0.3513& 0.9146& 0.5794& 0.5718& \cellcolor[HTML]{E5E5FF}0.2536& 2.5109& 1.8452& 2.5571\\
\textit{bicg}& 1.0109& 0.4150& 1.0771& \cellcolor[HTML]{E5E5FF}0.2463& 0.5229& 0.3147& 0.4738& \cellcolor[HTML]{E5E5FF}0.3212& 0.2711& 0.4205& 0.3480\\
\textit{gemm}& 1.1065& 0.5250& 1.0584& \cellcolor[HTML]{E5E5FF}0.4581& 0.5240& 0.5658& 0.6137& \cellcolor[HTML]{E5E5FF}0.9358& 0.6785& 0.5699& 0.7181\\
\textit{gesummv}& 0.6498& 0.6595& 0.9752& \cellcolor[HTML]{E5E5FF}0.5579& 0.4671& 0.4182& 0.3954& \cellcolor[HTML]{E5E5FF}0.3761& 0.4124& 0.4188& 0.4248\\
\textit{mvt}& 1.8678& 0.8715& 0.6403& \cellcolor[HTML]{E5E5FF}0.4420& 2.7230& 3.5088& 3.8933& \cellcolor[HTML]{E5E5FF}3.0261& 2.0321& 2.0427& 1.6964\\
\textit{md-knn}& 0.0217& 0.0216& 0.0064& \cellcolor[HTML]{E5E5FF}0.0064& 0.0246& 0.0224& 0.0306& \cellcolor[HTML]{E5E5FF}0.0526& 0.0250& 0.0232& 0.0215\\
\textit{spmv}& 0.4744& 1.0585& 0.8182& \cellcolor[HTML]{E5E5FF}0.0592& 0.3582& 1.1170& 1.3312& \cellcolor[HTML]{E5E5FF}0.5785& 1.0334& 1.7828& 1.3483\\
\textit{stencil2d}& 1.8879& 0.5093& 1.2729& \cellcolor[HTML]{E5E5FF}0.3764& 0.3378& 0.6214& 0.6710& \cellcolor[HTML]{E5E5FF}0.4525& 0.8438& 1.1927& 1.5266\\
\textit{stencil3d}& 1.7030& 0.6244& 1.5572& \cellcolor[HTML]{E5E5FF}0.1780& 0.4583& 0.6682& 0.3659& \cellcolor[HTML]{E5E5FF}0.4397& 0.7114& 0.6358& 0.7699\\
\textit{viterbi}& 0.1111& 0.1206& 0.2769& \cellcolor[HTML]{E5E5FF}0.0158& 0.2327& 0.2502& 0.2460& \cellcolor[HTML]{E5E5FF}0.1270& 0.1743& 0.1363& 0.1428\\
\textit{sha}& 0.3658& 0.1626& 0.7494& \cellcolor[HTML]{E5E5FF}0.7458& 0.2790& 0.1302& 0.2152& \cellcolor[HTML]{E5E5FF}0.2095& 0.3946& 0.2313& 0.2584\\
\textit{autocorr} & 0.0913& 0.0702& 0.0933& \cellcolor[HTML]{E5E5FF}0.1521& 0.0726& 0.0966& 0.1254& \cellcolor[HTML]{E5E5FF}0.0673& 0.0865& 0.0828& 0.0737\\ \midrule
\textbf{Avg}& \textbf{1} & \textbf{1.8492$\times$} & \textbf{1.5113$\times$} & \cellcolor[HTML]{E5E5FF}\textbf{4.4010$\times$} & \textbf{1.9408$\times$} & \textbf{1.8798$\times$} & \textbf{1.8042$\times$} & \cellcolor[HTML]{E5E5FF}\textbf{2.4461$\times$} & \textbf{1.4485$\times$} & \textbf{1.4221$\times$} & \textbf{1.3544$\times$} \\
\textbf{Geo Mean} & \textbf{1} & \textbf{1.6026$\times$} & \textbf{1.1404$\times$} & \cellcolor[HTML]{E5E5FF}\textbf{3.1335$\times$} & \textbf{1.5510$\times$} & \textbf{1.4415$\times$} & \textbf{1.3004$\times$} & \cellcolor[HTML]{E5E5FF}\textbf{1.6552$\times$} & \textbf{1.2223$\times$} & \textbf{1.2368$\times$} & \textbf{1.2046$\times$} \\ \bottomrule
\end{tabular}}
\end{table}

Lattice achieves favorable ADRS in smaller-scale designs by utilizing U-shaped Beta sampling \cite{DSE_lattice}.
This sampling approach is based on prior optimization experience that designs with either very high or very low parallelism often occupy opposite ends of the Pareto front (performance-optimal with higher hardware consumption versus resource-efficient with performance trade-offs).
U-shaped Beta sampling (BS) generates the initial configuration set $\varphi^0$ and derives the approximation of the Pareto front. For a sampling size \(N_0\), the space $\varphi^0$ comprises $n$ unique feature vectors $\varphi$, where each element of $\varphi$ is sampled probabilistically from a symmetric Beta distribution. The probability density function is defined over $0 \leq x \leq 1$ as:

\begin{equation}
\varphi^0(x) = \frac{x^{\alpha-1}(1-x)^{\alpha-1}}{\mathrm{B}(\alpha)},
\end{equation}

\begin{equation}
\mathrm{B}(\alpha) = \int_{0}^{1} x^{\alpha-1}(1-x)^{\alpha-1} dx.
\end{equation}

where $\mathrm{B}(\alpha)$ denotes the Beta function. This U-shaped distribution (with $\alpha < 1$) assigns higher probability density to boundary values of $x$. This sampling strategy ensures that the initial feature vectors $\varphi^0$ contain extreme parameter values with high likelihood, intentionally reserving intermediate feature combinations for exploration during subsequent refinement stages.

The uniform random sampling (RS) strategy generates independent feature vectors $\varphi^0$ by selecting each parameter $\varphi_j$ (for $j=1,...,d$ dimensions) from a uniform distribution.

Latin Hypercube Sampling (LHS) enforces stratified spatial coverage in $d$-dimensional space through orthogonal permutation:

\begin{equation}
\varphi^0_{ij} = \frac{\pi_j(i) - u_{ij}}{n}, \quad u_{ij} \sim \mathcal{U}(0,1)
\end{equation}

where $\pi_j(\cdot)$ denotes a random permutation function for the $j$-th dimension, and $u_{ij} \sim \mathcal{U}(0,1)$ denotes uniformly distributed stochastic offsets within hypercube cells.

\textbf{Seed Directive Generation} approach (\textit{Warm-Start}) integrates a generative model $\pi_\theta(\cdot)$, a prompt $\mathcal{P}$ encoding task constraints, and prior knowledge of hardware optimization $\mathcal{K}$. Its sampling mechanism is governed by:

\begin{equation}
\varphi^0_i \overset{\text{i.i.d.}}{\sim} \pi_\theta(\cdot \mid \mathcal{P}, \mathcal{K}) 
\end{equation}

\vspace{5pt} 

where $\varphi_i$ denotes a $d$-dimensional feature vector. LLMs achieve superior efficiency by integrating domain knowledge and constrained optimization principles, outperforming other sampling baselines.

\subsection{Analysis of LLM Performance in DSE}

In this section, we provide additional experimental results to elucidate LLM performance during the \textit{Preprocessing}, \textit{Warm-Start}, and \textit{Adaptive Optimization} phases of DSE tasks. We also conduct extensive comparisons with current state-of-the-art general-purpose LLMs.

\begin{wrapfigure}{r}{0.5\textwidth}
    \centering
    \includegraphics[width=\linewidth]{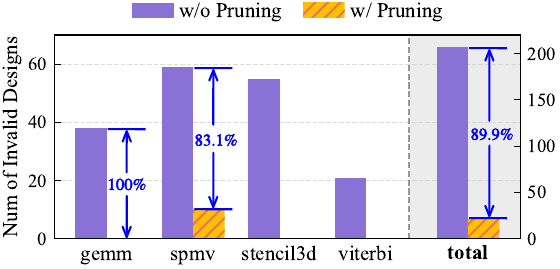}
    \caption{Comparison of invalid design counts with and without \textit{Feature-Driven Pruning}}
    \label{fig:prune}
\end{wrapfigure}

We provide supplementary data for Figure \ref{fig:ablation} in our ablation study. To validate the effectiveness of our \textbf{Feature-Driven Pruning} method, we compared the design space dimension in the \textit{Adaptive Optimization} phase with and without pruning. We recorded the number of invalid designs that failed due to synthesis timeout or failure caused by scheduling excessive parallelism directives, as shown in Figure \ref{fig:prune}. The experimental results demonstrate that for benchmarks with larger design spaces and nested loops with high trip counts \textit{(gemm, spmv, stencil3d, viterbi)}, the \textit{w/o Pruning} approach encountered numerous invalid designs. This occurs because existing LLMs struggle to capture the semantics of complex HLS DSE tasks and lack specialized hardware optimization knowledge. Consequently, they fail to establish awareness of QoR feedback during exploration, becoming lost in the vast design space. In contrast, constraining exploration to the pruned design space allows LLMs to explore directive parallelism within a restricted feasible region. This approach prevents wasted computational resources resulting from aggressive parallelism allocation or redundant directive configurations that lead to failed QoR evaluations of candidates. Moreover, our experimental results indicate that this mild pruning strategy not only significantly reduces invalid designs but also maintains excellent DSE performance, as evidenced by further improvements in Figure \ref{fig:ablation}.

\vspace{10pt}
\begin{figure*}[h]
	\centering 
 \includegraphics[width=14cm]{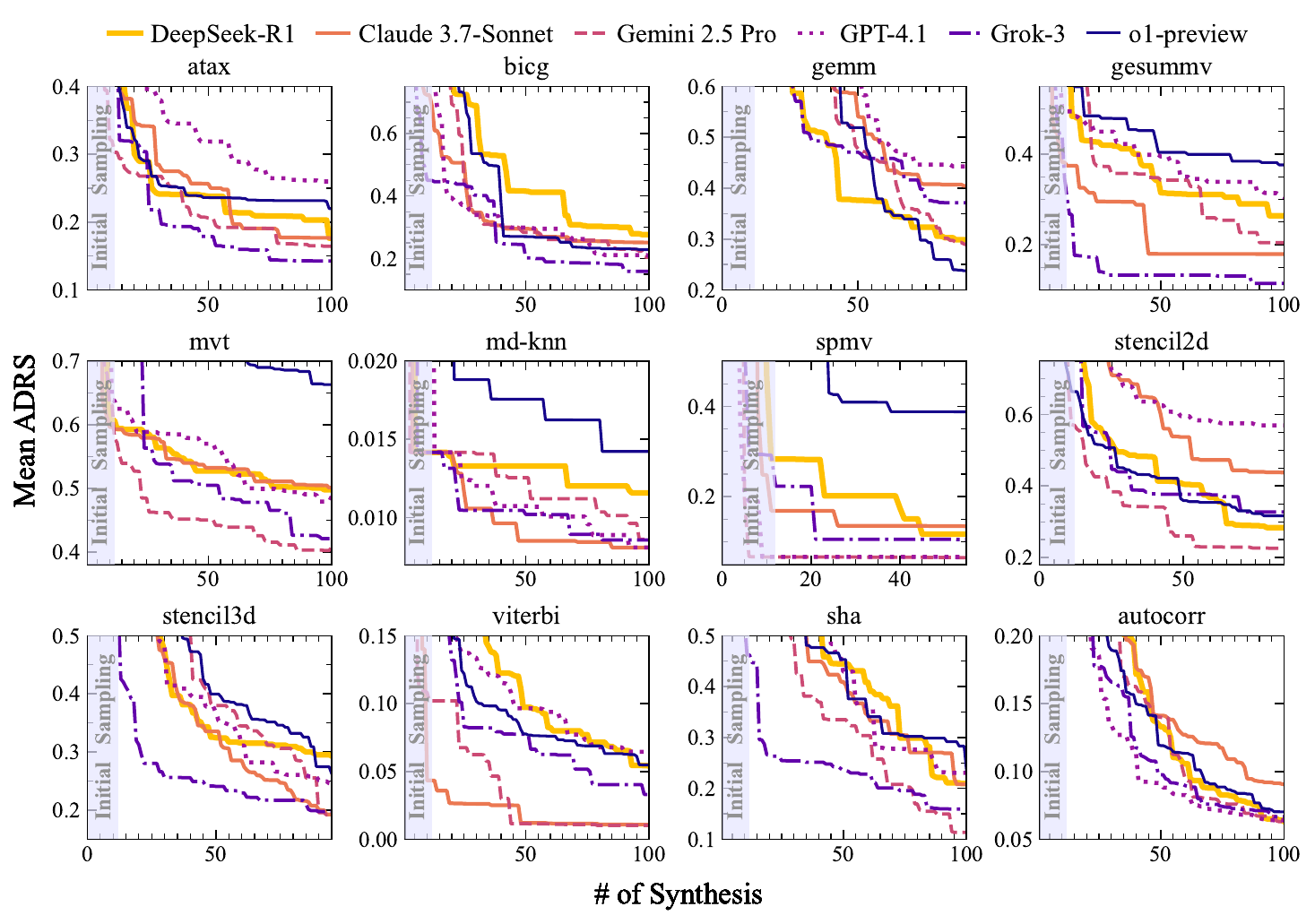} 
	\caption{Comparison of ADRS among MOEA/D-based DSE method under different LLMs.}
	\label{moead_adrs}
 \vspace{5pt}
\end{figure*}

\vspace{10pt}

We validated our approach across multiple inference LLMs, comparing various general-purpose LLMs including \textit{DeepSeek-R1}, \textit{Claude 3.7-Sonnet}, \textit{Gemini 2.5 Pro}, \textit{GPT-4.1}, \textit{Grok-3}, and \textit{o1-preview} for DSE. As training scales expand and methodologies evolve, LLMs show increasing promise for driving effective DSE.
Importantly, we believe that the vast HLS design space prevents LLMs from memorizing optimal directive configurations during pre-training. Instead, LLMs allocate these configurations based on learned hardware optimization principles, preserving the flexibility and generalization of iDSE.

\vspace{10pt}

\begin{figure*}[h]
	\centering 
 \includegraphics[width=14cm]{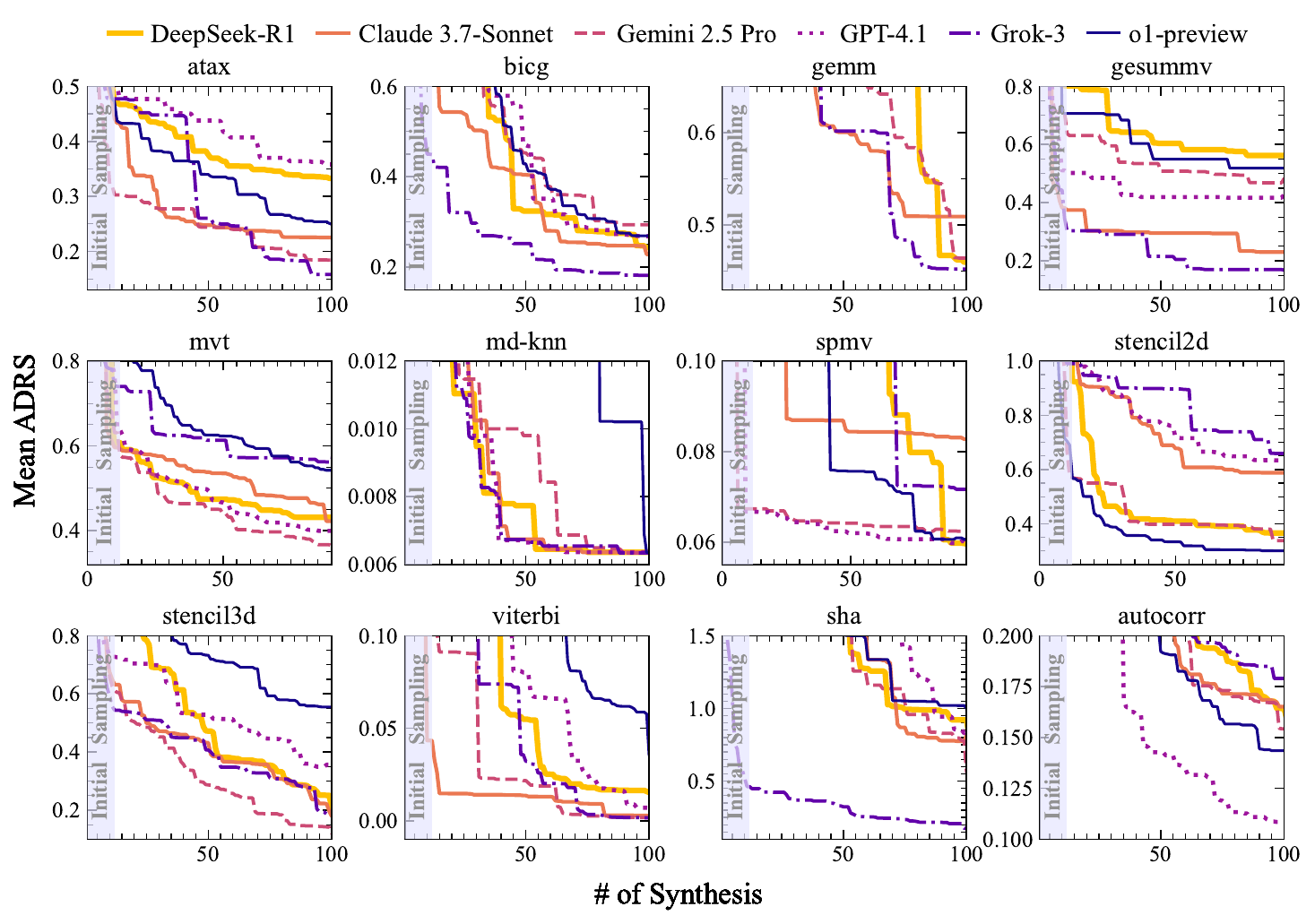} 
 \vspace{-15pt}
	\caption{Comparison of ADRS among NSGA-\uppercase\expandafter{\romannumeral2}-based DSE method under different LLMs.}
	\label{nsga_adrs}
\end{figure*}

\vspace{15pt}

We compared different general-purpose LLMs for initial sampling, followed by EA-based DSE methods implemented with MOEA/D (Figure \ref{moead_adrs}) and NSGA-II (Figure \ref{nsga_adrs}) for searching new directive configurations. Our experiments revealed significant variations in both convergence speed and final optimization quality across different LLMs. Notably, we found that no single model consistently outperformed others across all benchmarks, indicating that LLMs exhibit different optimization reasoning patterns depending on the HLS design structure and design space dimension.

\begin{figure*}[h]
	\centering 
 \includegraphics[width=14cm]{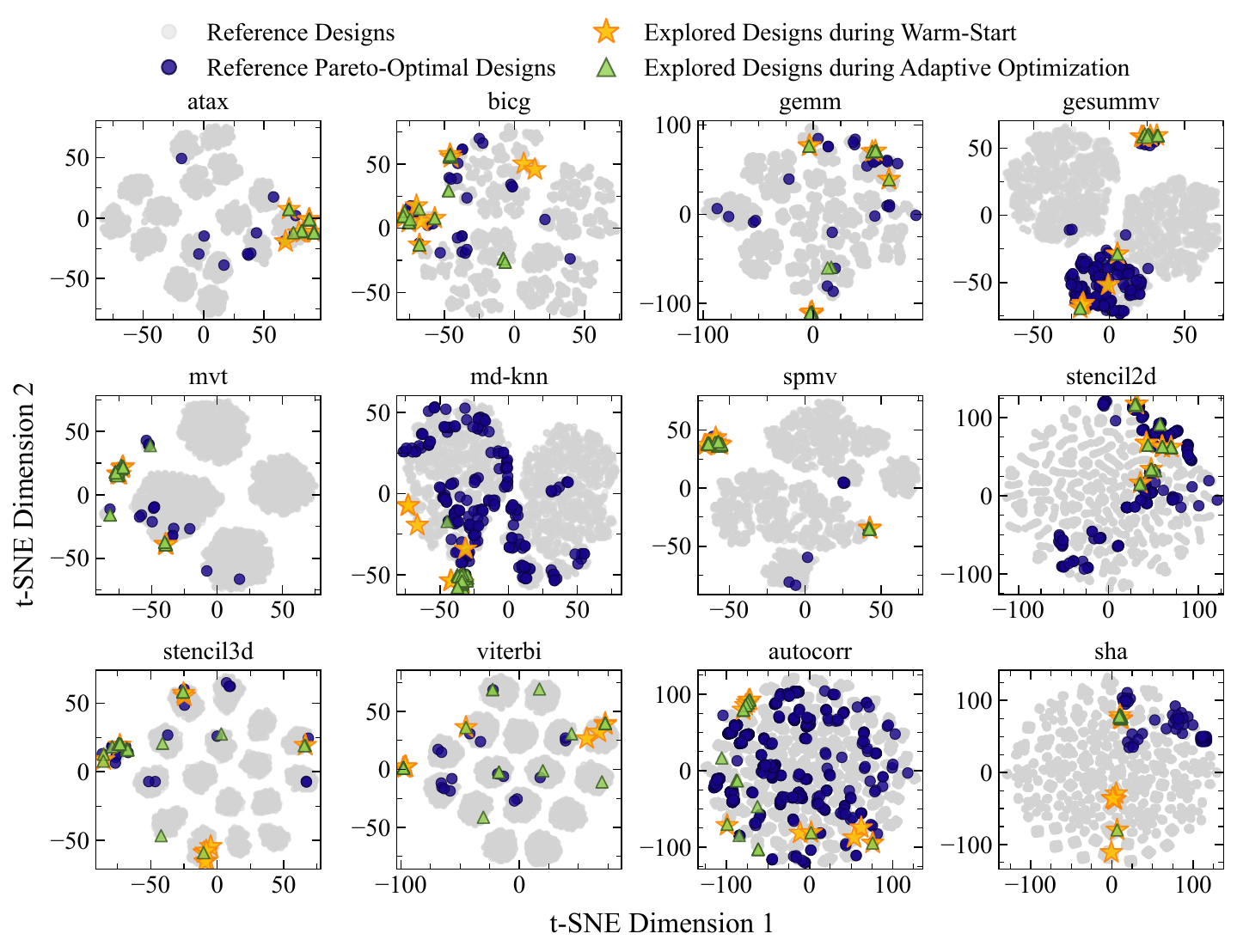} 
	\caption{t-SNE visualization of reference Pareto-optimal and explored optimization directives in \textit{Warm-Start} and \textit{Adaptive Optimization} stages}
	\label{tsne}
\end{figure*}

Figure \ref{tsne} compares the t-SNE visualization of feature vectors from Pareto-optimal directive configurations sampled through exhaustive exploration (approximately 10,000 explored designs per benchmark) against those identified by our method during both initial sampling and search phases. t-SNE is a dimensionality reduction technique that represents high-dimensional data as points in a two-dimensional space, where proximity between points indicates similarity in the data, while greater distances suggest dissimilarity.
The visualization reveals that the design space formed by HLS optimization directives along with their parameters and combinations is exponentially large, making feature differentiation challenging and resulting in irregular visualization patterns. Furthermore, most HLS designs exhibit clustering characteristics among Pareto-optimal directive configurations. Our \textit{Warm-Start} strategy effectively occupies these optimal design regions, thereby guiding subsequent \textit{Adaptive Optimization} phases toward broader coverage.
Comparative analysis with Figure \ref{experiment1} demonstrates that while our method may not comprehensively cover the extensive Pareto-optimal designs identified through exhaustive exploration under limited search budgets, it successfully constructs impressive Pareto fronts in the objective space (latency-resource utilization). This approach effectively satisfies design requirements while significantly reducing exploration overhead.

\subsection{Information of Assets}
\label{app:asset}

We provide all results of our experiments in our double-blind review repository, and we plan to upload our complete code for public after publication.

We present the information of assets as below:

\begin{enumerate}[leftmargin=*]

\item Code

\begin{itemize}
\item HGBO-DSE \cite{hgbo}
\begin{itemize}
\item License: MIT license.
\item URL: https://github.com/hzkuang/HGBO-DSE
\end{itemize}
\end{itemize}

\begin{itemize}
\item Lattice \cite{DSE_lattice}
\begin{itemize} 
\item License: Available online. 
\item URL: http://www.inf.usi.ch/phd/ferretti/lattice-traversing-DSE.html
\end{itemize}
\end{itemize}

\item Dataset

\begin{itemize}

\item PolyBench \cite{bench_polybench}
\begin{itemize}
\item License: Ohio State University Software Distribution License.
\item URL: http://polybench.sf.net.
\end{itemize}

\item CHStone \cite{bench_chstone}
\begin{itemize} 
\item License: Available online.
\item URL: https://github.com/ferrandi/CHStone
\end{itemize}

\item MachSuite \cite{bench_machsuite}
\begin{itemize} 
\item License: MachSuite BSD-3 license.
\item URL: https://github.com/breagen/MachSuite
\end{itemize}

\end{itemize}

\end{enumerate}

\section{Limitations and Future Work}
\label{app:limit}

One limitation of our work is that LLMs cannot identify all ground-truth Pareto-optimal designs under finite search budgets, while exhaustive exploration remains impractical due to the vast design space. However, our experimental results demonstrate that iDSE significantly outperforms heuristic-based DSE methods, discovering Pareto fronts that, although potentially less diverse, are sufficiently impressive to satisfy multifaceted optimization preferences. Relaxing the maximum iteration limit would allow convergence toward a wider spectrum of Pareto-optimal designs at increased time cost. However, the effectiveness of LLMs may diminish with expanding exploration scales due to their limitations in handling long contexts. Our observations indicate that representative initial sampling designs generated by the LLM effectively enhance subsequent traditional evolutionary algorithms. Therefore, our method could be further improved by focusing on providing broader initial sampling designs combined with traditional DSE methods, thereby leveraging their established advantages in long-term hardware optimization.

Additionally, despite subsequent refinement, a modest number of low-quality directive configurations in the initial sampling still result in wasted evaluation time of the vendor HLS tool. We also observed that different general-purpose LLMs exhibit varying performance across benchmarks, with no single model consistently outperforming others, potentially complicating model selection for designers. A promising solution involves developing specialized models through targeted training, thus enhancing the scalability of the method, which we consider as an improvement direction for future work in iDSE.
Furthermore, our iDSE framework can also serve as a data augmentation method for collecting high-quality datasets in DSE to drive future model training.
Despite these limitations, we believe that iDSE represents pioneering work which provides valuable insights into the future of processor-level design space exploration, motivates efforts to apply generative AI to hardware optimization, and pushes the boundary of electronic design automation.

\end{document}